\documentclass[prd,showpacs,amsmath,amssymb,aps,floats]{revtex4}%
\usepackage{epsfig}
\usepackage{dcolumn}
\usepackage{bm}
\usepackage{amsmath}
\usepackage{amsfonts}
\usepackage{amssymb}
\def\nnd{\end{document}}

\def\be{\begin{equation}}
\def\ee{\end{equation}}

\newcommand{\bea}{\begin{eqnarray}}
\newcommand{\eea}{\end{eqnarray}}
\newcommand{\bwt}{\begin{widetext}}
\newcommand{\ewt}{\end{widetext}}

\def\u
\def\hZ{\widehat Z}

\def\eed{\end{document}}

\def\m_z{m_{\textrm {Z}}}

\renewcommand{\d}{\rm{d}}
\renewcommand{\u}{\rm{u}}
\newcommand{\q}{\rm{q}}
\newcommand{\s}{{\rm{s}}}

\def\be{\beta}

\makeatother
\begin{document}
\title{The pseudoscalar glueball in a chiral Lagrangian model with instanton effect}
\author{Song He$^{1}$}
\author{Mei Huang$^{1,2}$}
\author{Qi-Shu Yan$^{3}$}
\affiliation{$^{1}$ Institute of High Energy Physics, Chinese Academy of Sciences, Beijing
100039, China }
\affiliation{$^{2}$ Theoretical Physics Center for Science Facilities, Chinese Academy of Sciences, Beijing, China}
\affiliation{$^{3}$ Department of Physics, University of Toronto, Toronto, Canada }

\begin{abstract}
We study the pseudoscalar glueball candidates in a chiral effective Lagrangian model proposed by 't hooft, motived by taking into account the instanton effects, which can describe not only the chiral symmetry breaking, but also the solution of $U_A(1)$. We study the parameter space allowed by constraints from vacuum conditions and unitary bounds. By considering two scenarios in $0^{++}$ sector, we find that parameter space which can accommodate the $0^{-+}$ sector is sensitive to the conditions in $0^{++}$ sector. From our analysis, it is found that three $\eta$ states, i.e. $\eta(1295)$, $\eta(1405)$, $\eta(1475)$, can be glueball candidates if we assume that the lightest $0^{++}$ glueball has a mass $1710$ MeV. While there is no $0^{-+}$ glueball candidate found in experiments if we assume that the lightest $0^{++}$ glueball has a mass $660$ MeV.
\end{abstract}

\maketitle


\renewcommand{\thefootnote}{\arabic{footnote}} \setcounter{footnote}{0}%

\setcounter{equation}{0}
\renewcommand{\theequation}{\arabic{section}.\arabic{equation}}%

\section{Introduction}

\begin{figure}[t]
\centerline{
\epsfxsize=12 cm \epsfysize=12 cm \epsfbox{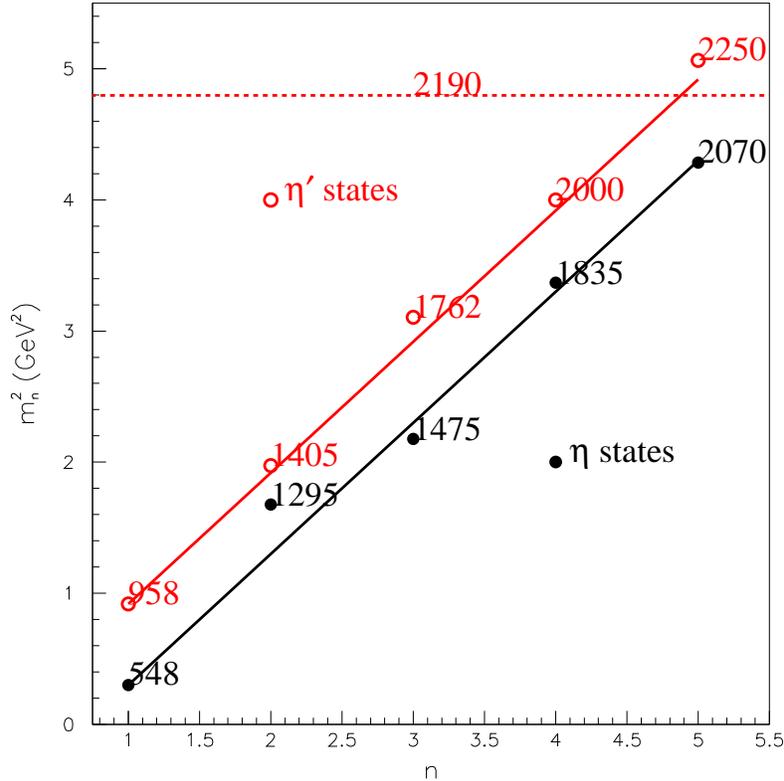}
}
\caption{\textit{The $\eta$ and $\eta^\prime$ spectra on Regge lines. The slope of these two lines is taken as $1$ GeV$^2$. }}
\label{fig-regge}
\end{figure}

Currently, as provided in PDG \cite{Amsler:2008zz} and other literature \cite{Godfrey:1998pd,Amsler:2004ps,Klempt:2007cp}, there are about $11$ $0^{-+}$ particles identified in total, which are tabulated in Table \ref{table1}. It is urgent to know which are normal $q{\bar q}$ states and which are exotic states. For a ground $q{\bar q}$ state, in the quasiclassic picture, the quark and antiquark are attached to a string with the tension $\sigma$. The mass squared of its higher excitation states can be described as $m^2_n = 4 \pi \sigma n$. This linear n dependence can be used to classify these states properly and find out exotic states. Therefore, it is useful and helpful to put these states on the Regge trajectories, as shown in Fig. (\ref{fig-regge}) where we assume that $4 \pi \sigma = 1$ which is consistent with the analysis presented in \cite{Anisovich:2000kx}. We find that these states can be well-organized by the radial excitations of $\eta$ and $\eta^\prime$, except the state $\eta(2190)$, which is proposed as a glueball candidate in \cite{Bugg:1999jc}.

Let's comment on these resonances briefly: the first two are well-known $\eta$ and $\eta^\prime$. For the following three, the $\eta(1295)$ is believed to be the first Regge excitation of $\eta$ or even doesn't exist \cite{Klempt:2007cp}. The $\eta(1405)$ is possibly a glueball while the $\eta(1475)$ is believed to be the Regge excitations of ${\bar s} s$ \cite{Amsler:2008zz,Amsler:2004ps}. Or there is no split at all and there is just the radial excitation of $\eta$. The $\eta(1760)$ is the radial excitations for $\eta$, while the $X(1835)$ is assumed to be the first Regge excitation of $\eta^\prime$ \cite{Klempt:2007cp}. The $\eta(2070)$ or the $\eta(2190)$ can be interpreted as a pseudoscalar glueball \cite{Bugg:1999jc} or just the second Regge excitation of $\eta^\prime$ \cite{Klempt:2007cp}. The last $\eta(2250)$ should also be the radial excitation of $\eta$. Interested readers can refer to \cite{Godfrey:1998pd,Amsler:2004ps,Klempt:2007cp} for further reading and more details. 
 
\begin{table}[th]
\begin{center}%
\begin{tabular}
[c]{|c|c|c|c|c|c|c|c|c|c|c|c|}\hline
&  &  &  &  &  &  & & & &&\\
& $\eta$ & $\eta^\prime$ & ($\eta$) & $\eta$ &$\eta$ &  $\eta$ & $X$ & $\eta$& $\eta$ & $\eta$ &  $\eta$ \\\hline
&  &  &  &  & & & & &&&\\
$M_\eta$ & $548$ & $958$ & $(1295)$ & $1405$ &  $1475$ &
$1760$ & $1835$ & $2000$ & $2070$ & $2190$ & $2250$ \\
&  &  &  &  &  & & & & && \\\hline
\end{tabular}
\end{center}
\caption{ The $11$ spectra of $\eta$ mesons (The unit is in
MeV) .}%
\label{table1}%
\end{table}

Another interesting and long-time quest is for glueball candidates. Glueball states are the consequence of the non-Abelian nature of QCD gauge theory. The first pseudoscalar glueball candidate was found in Mark II and Crystal Ball Collaboration and was dubbed as $\iota(1440)$ or $G(1440)$. Nowadays,  this particle is named as $\eta(1405)$ by PDG. Currently, it is widely believed that glueballs can be produced in the $J/\Psi$ radiative decay via the $\gamma g g$ channel, where the final two gluons have a considerable probability to form a glueball bound state. Meanwhile, a pure glueball state is forbidden to decay into a photon pair or a lepton pair at the leading order ( which is known as the rule of thumb OZI suppression ), therefore its partial decay widths to these channels should be suppressed and should be much smaller than those of a $q{\bar q}$ state. Such a feature can be used to distinguish hardonic and glueball states. For instance, by study the productions and decays of those $\eta$ states, several groups \cite{LZC07,Cheng:2008ss} come to the conclusion that $\eta(1405)$ is a good glueball candidates. At the same time, there are other states, like $\eta(1760)$ \cite{Wu:2000yt} and $X(1835)$ \cite{KM05}, which could also be a glueball state. The consensus on the question which is glueball among these $0^{-+}$ states has not been achieved yet.

It is well-known a realistic glueball state is  a mixture of constituent gluons and quarks. A large mixing prohibits us to isolate a hadronic state from a glueball state via either productions or decays. Therefore it seems difficult to have a definite and positive answer to the question on the $\eta$ spectra: which spectrum might be the glueball candidates? Then it is urgent to have a trustable and systematic method to describe the glueball and meson mixing and their interaction, while the effective Lagrangian method is a good one for such a purpose.

The effective Lagrangian method is a useful method for hadron physics. The advantage of this method lies in the fact that the global symmetries of QCD can be realized in a convenient way. It is  natural to use this method to describe the psuedo-scalar glueball 
candidates. In principle, the effective Lagrangian should be derived from QCD, as shown by Cornwall and Soni \cite{Cornwall:1984pa} for pure glueball sector, or as demonstrated by Gasser and Leutwyler \cite{Gasser:1983yg,Gasser:1984gg} for the case when only Goldstone mesons are concerned. It is well-known it is difficult to derive the effective Lagrangian from QCD for the general cases, and one popular way is to construct the lowest order operators in the Lagrangian by using the global symmetries of QCD as a guiding principle \cite{Weinberg:1978kz}. 

It is well-known that pseudoscalars play a special role in the low energy hadronic QCD physics: they are the Goldstone particles which tightly relate with the spontaneous chiral symmetry breaking. In QCD quark model with three flavors, it is observed that the quark fields possess a $U(3)_R \times U(3)_L$ chiral symmetry, which is larger than the $SU(3)_R \times SU(3)_L$. However, if the chiral symmetry breaking is a correct concept to the pseudoscalar sector, it is expected that the lowest Goldstone particles should form a nonet for the symmetry breaking of $U(3)_R \times U(3)_L \to U(3)_D$, instead of an octet observed in experiments which is consistent with the chiral symmetry breaking $SU(3)_R \times SU(3)_L \to SU(3)_D$. This is the famous $U_A(1)$ problem for QCD quark model. A workable model should describe not only the chiral symmetry breaking, but also  the solution of $U_A(1)$.

One of the early efforts is to extend the large $N_c$ effective chiral Lagrangian \cite{Witten:1979vv,Veneziano:1979ec} proposed to solve the $U_A(1)$ problem by including the sector for the glueball candidates, like those authors of \cite{Schechter:1980ak,Rosenzweig:1981cu,Rosenzweig:1982cb} did by coupling the scalar glueball field to the logarithmic determinant of chiral scalar field. Another way is to extend the chiral Lagrangian of meson states by including a glueball field, as shown in references \cite{Giacosa:2005zt}. The drawbacks of these models lie in the fact that most terms in the models are put by hand or QCD global symmetries are not realized in a transparent way. A model motived from QCD might be a better one.
 
An alternative chiral symmetry model, proposed by 't Hooft \cite{'tHooft:1986nc}, can solve the $U_A(1)$ problem by using the instanton effects \cite{'tHooft:1976fv}. In response to the criticism from \cite{Christos:1984tu}, 't Hooft showed that indeed by including a spurion field (which will be interpreted as glueball state in this paper) coupling to the determinant of chiral scalar field, not the logarithmic one, the effective Lagrangian is consistent with the global chiral symmetry and the $U_A(1)$ symmetry of QCD. Meanwhile, in this form, the solution of the $U_A(1)$ problem can be realised in a spontaneous way, although it is not necessary to be. The $U_A(1)$ problem is neatly solved in this model. The $\eta^0$ corresponds to the $\eta^\prime$ observed in experiments. The determinant term induced from instanton effects can split the octet and singlet of the nonet and can make $\eta^0$ heavier than $\eta^8$. Therefore this effective Lagrangian is well justified, and we will use this effective Lagrangian as a framework to describe the mixing and interactions of pseudoscalar glueballs and $\eta$ mesons.

In this work, we study the vacuum structure and the pseudoscalar spectra ($\eta$, $\eta^\prime$, and psuedo-scalar glueball states) and their mixing. By making a connection between the FKS formula and the effective model, we find some correspondences between the effective parameters and the quantities computable in lattice QCD.  We also investigate the theoretical uncertainties and the lattice computations of the model parameters. We explore the parameter space of the model from the constraints of the $0^{++}$ and $0^{-+}$ sector and study some phenomenologies related to the pseudoscalar glueball candidates. It is found that three  $\eta$ states, i.e. $\eta(1295)$, $\eta(1405)$, $\eta(1475)$, can be glueball candidates if we assume that the lightest $0^{++}$ glueball has a mass $1710$ MeV. While there is no $0^{-+}$ glueball candidate if we assume that the lightest $0^{++}$ glueball has a mass $660$ MeV.

The paper will be organized as follows. The 4D linear chiral Lagrangian with a term induced by instanton effect is introduced in Section II. The mass matrices for the $0^{++}$ and $0^{-+}$ sector  are given in Section III. The theoretical uncertainties and lattice results in determining $k$ are explored in Section IV. The effective mass of gluon is also explored in this Section.
We analyze the parameter space determined from vacuum conditions and $0^{++}$ and $0^{-+}$ scenarios in Section V. The phenomenological consequence of pseudoscalar is also studied in this Section. We end this paper with discussions and conclusions.

\section{Pseudo-scalar glueball in the 4D linear/nonlinear $\sigma$ model}
The linear $\sigma$ model has been utilized to accommodate the chiral symmetry breaking in QCD since 1960's \cite{GellMann:1960np}. In order to accommodate the glueball candidates, we must extend the ordinary $\sigma$ model by including a glueball field. The effective Lagrangian is provided in \cite{'tHooft:1986nc} and can be formulated into two parts, the first part is symmetric and the other part includes explicit symmetry breaking terms. The effective Lagrangian reads as
\bea
\cal{L} &=& \cal{L}_S + \cal{L}_B \,.
\label{totl}
\eea
The symmetry part can be put as
\bea
\label{model}
\cal{ L}_S &=& Tr[\partial X^\dagger \cdot \partial X]  - m_X^2 Tr[X^\dagger \,\, X] + \partial Y^* \cdot  \partial Y - m_Y^2 Y^* \,\, Y\\ \nonumber
&&-\frac{\lambda_{XY} }{2} Tr[X^{\dagger} X] Y^* Y  - \frac{\lambda_{1,X}}{4} Tr[X^\dagger X X^\dagger X]   - \frac{\lambda_{2,X}}{4} Tr[X^\dagger X]^2  \\ \nonumber 
&& -\frac{\lambda_Y}{4} (Y^* Y)^2 - \left ( \frac{ k} {6} \,\, Y \,\, det(X) + h.c. \right ) + \cdots \,, 
\eea
where $X$ is a $3\times 3$ bi-fundamental field which is to describe the scalar mesons (both the $0^{++}$ nonet and the $0^{-+}$ nonet) and Y is a complex scalar (also called "spurion field") which is supposed to describe the pure gluon-bound states here.

The terms which explicitly break symmetry breaking are given as 
\bea
\cal{ L}_B &=& \left ( \frac{1}{2} Tr[ B \cdot X] + h.c. \right) + \left (\frac{1}{2} D\,\, Y + h.c. \right ) + \cdots \,,
\eea
which are introduced to describe the mass pattern of Goldstone particles due to the explicit chiral symmetry breaking. In the chiral Lagrangian, the parameter $B$ is a matrix which is proportional to the quark mass and is parameterized as $B = B_0 v_X m_q$, where $m_q$ is a mass matrix of quarks. Similar to the $B$ parameter, the $D$ parameter should be proportional to the effective gluon mass and can be parameterized as $D=D_0 \frac{v_Y}{3} m_g^{eff}$.

It is helpful to count the mass dimension of each variable and quantity in the Lagrangian. The fields $X$ and $Y$ have mass dimension [GeV]. The parameters $B$ and $D$ have mass dimension [GeV]$^3$. The parameter $k$ is dimensionless. In this paper, for the sake of simplicity, we assume these parameters are real.

The Lagrangian $\cal{L}_S$ is invariant under the chiral flavor symmetry  $SU(3)_L \times SU(3)_R$ transformation:
\bea
X \rightarrow X^\prime = U_R \,\, X \,\, U_L^\dagger \,,
\eea
where $U_R$ and $U_L$ are group elements of the $SU(3)_L\times SU(3)_R$ symmetry, which can be parameterized as
$exp (i \theta_R^a T^a)$ and $exp(i \theta_L^a T^a)$, respectively.
$T^a$ with $a=1, \cdots, 8$ are generators of $SU(3)$ Lie group, and $\theta^a$ are parameters of the group space. The Lagrangian $\cal{L}_S$ is also invariant under the $U_A(1)$ transformation:
\bea
X & \rightarrow X^\prime = & exp(  i \theta_A T^0) X\,, \\ \nonumber
Y & \rightarrow Y^\prime = & exp(- i~ 3~ \theta_A) Y\,.
\eea
Where $T^0$ is the $3\times 3$ identity matrix and $\theta_A$ is a  phase angle.

The chiral symmetry $SU(3)_L \times SU(3)_R$ of the $\cal{L}_S$ is broken by the nonvanishing vacuum expectation value $v_X$ of $X$ field to the diagonal $SU(3)_D$. Meanwhile, the $U_A(1)$ symmetry is also broken by the nonvanishing vacuum expectation value $v_Y$ of $Y$ field. Therefore, at low energy region, the relevant degree of freedom is Goldstone bosons. Therefore we can parameterize relevant Goldstone fields in a nonlinear way as
\bea
X &\to &v_X exp( i \frac{ \eta^a}{v_X} T^a) \,,\\\nonumber
Y &\to & \frac{ v_Y }{ 3} exp(i \frac{3}{v_Y} a) \,,
\label{goldstone}
\eea
where $\eta^a$ denote the Goldstone particles which break the  $SU(3)_L \times SU(3)_R$ to $SU(3)_D$, $a$ denotes the Goldstone particle which breaks the $U_A(1)$ symmetry. 

The vacuum expectation values in the parameterization of Goldstone field have two functions: 1) make the field $\eta^a$ and $a$ having the right dimension in 4D; 2) make the kinetic term of all field being normalized to one. The advantage of the nonlinear parameterization is that the quartic terms in the Higgs potential do not affect the Goldstone interactions and the model can be easily compared with the chiral Lagrangian method.

We will concentrate on the study of the effects of the instanton induced determinant term, and hereafter we drop the term proportional to $\lambda_{XY}$, which also describes the interaction between mesons and glueballs. For the sake of convenience, here the vacuum is  assumed to have the following form 
\bea
\langle X \rangle &=&  v_X ~ I_{3\times 3}\,,\\
\langle Y \rangle &=&  \frac{v_Y}{3} \,.
\eea
Here we already assume that $v_Y$ is real and has no complex angle.

To have the spontaneous symmetry breaking, it is required that the vacuum should be stable. Therefore the parameters in the potential  should satisfy the following two constraints
\bea
-3 B + 6 m_X^2 v_X + \frac{k}{3} v_Y v_X^2 +3  \lambda_X v_X^3  &=& 0\,,\label{cond1}\\ 
- D + \frac{2}{3} m_Y^2 v_Y +\frac{k}{3} v_X^3 + \frac{1}{27} \lambda_Y v_Y^3 &=& 0\,,\label{cond2}
\eea
where $\lambda_X = \lambda^X_1 + \lambda_2^X$.

\setcounter{equation}{0}
\renewcommand{\theequation}{\arabic{section}.\arabic{equation}}%
\section{Mass matrices of $0^{++}$ and $0^{-+}$ sectors}

There are 8 free parameters in the model. In order to reduce the number of free parameters and to simplify the analysis, we can use the information from the $0^{++}$ sector to constrain the parameter space.

For the sake of simplicity, here we only consider the mixing between
the pure $0^{++}$ glueball (the modulus of complex $Y$ field minus the vacuum expectation value $v_Y$) and the singlet $\sigma^0$ field. Then the mass matrix of $0^{++}$ scalars, which determines the masses of glueball and $q{\bar q}$ states, is given as
\begin{eqnarray}
\label{massm}
M_h^2 = \left ( \begin{array}{cc}
M_{XX}^2 & M_{XY}^2   \\
&\\
M_{XY}^2 & M_{YY}^2  \\
\end{array} \right )\,.
\end{eqnarray} 
While the mass matrix elements are given as
\bea
M_{XX}^2 &=&6 m_X^2+9 \lambda_X  v_X^2+\frac{2}{3} k\, v_Y\, v_X\,, \\
M_{YY}^2 &=& 2 m_Y^2 + \frac{\lambda_Y}{3} v_Y^2\,,\\
M_{XY}^2 &=& k\, v_X^2\,.
\eea
We can use the physics mass of $0^{++}$ glueball and $q{\bar q}$ states and the mixing angle to represent these elements as
\bea
M_{q{\bar q}}^2 \, cos^2\alpha + M_{gg}^2 \, sin^2\alpha & = & M_{XX}^2 \,, \label{rep1}\\
M_{q{\bar q}}^2 \, sin^2\alpha + M_{gg}^2 \, cos^2\alpha & = & M_{YY}^2 \,,\\
\frac{1}{2} \left ( M_{gg} ^2 - M_{q{\bar q}}^2 \right ) sin\,2 \alpha  & = & M_{XY}^2 \,. \label{rep2}
\eea
Then by using Eqs. (\ref{cond1}-\ref{cond2}) and Eqs. (\ref{rep1}-\ref{rep2}) with six inputs ( the experimental determined values of $v_X$, the mass of $0^{++}$ glueball candidate $M_{gg}$, the mass of $q{\bar q}$ meson state $M_{q{\bar q}}$, the assumed mixing angle $\alpha$, the assumed value of $v_Y$  and $m_g^{eff}$ ), we can solve out the variables $m_X^2$, $m_Y^2$, $\lambda_X$, $\lambda_Y$, and $k$.

In order to analyze the symmetry breaking pattern, we choose two scenarios to scan the parameter space of $\lambda_X$  and $\lambda_Y$. The masses of scalars are chosen as given in Table \ref{table-cases}. Scenario one is motivated from the possibility that $\sigma(660)$ might be a glueball rich state \cite{Minkowski:1998mf,Vento:2004xx}. Scenario two is motivated from that $f_0(1710)$ is a good glueball candidate, as proposed by \cite{Cheng:2006hu} and supported by the lattice computation \cite{Chen:2005mg}. It is found this scenario is also favored in the recent study under the context of AdS/QCD \cite{Wang:2009wx}. We assume the mixing angle is a free parameter for these two cases.
\begin{table}[th]
\begin{center}%
\begin{tabular}
[c]{|c|c|c|}\hline
                  & & \\
                  & $0^{++}_{gg}$ & $ 0^{++}_{ q {\bar q}} $   \\
                  & & \\ \hline
                  & & \\
Scenario 1 & $0.66$ & $0.98$ \\
                  & & \\\hline
                  & & \\
Scenario 2 & $1.71$ & $1.37$\\
                  & & \\ \hline
\end{tabular}
\end{center}
\caption{ Two scenarios in $0^{++}$ sector are considered for the study of symmetry breaking pattern and for the constraint on the parameter space. }%
\label{table-cases}%
\end{table}
By requiring that the potential should have a bottom when $X\to \infty$ and $Y \to \infty$, we can impose the following two conditions $\lambda_X>0$ and $\lambda_Y>0$. At the same time, we require that the value of $\lambda_X$ and $\lambda_Y$ should not be too large to violate the unitary conditions then we impose two conditions $\lambda_X<40$ and $\lambda_Y<40$.

From the solutions to the five equations, we observe that there are four patterns of symmetry breaking in the parameter space: 1) Symmetries are broken by the conditions $m_X^2 <0$ and $m_Y^2 < 0$ with $k > 0$, which corresponds to the case where these  symmetries are broken independently to each other. 2) Symmetries are broken by the conditions $m_X^2 < 0$ and $m_Y^2 > 0$ with $k > 0$, which corresponds to the case where the chiral symmetry breaking triggers the $U_A(1)$ symmetry breaking. 3) Symmetries are broken by the conditions $m_X^2 > 0$ and $m_Y^2 < 0$ with $k > 0$, which corresponds to the case where the $U_A(1)$ symmetry breaking triggers the chiral symmetry breaking. 4) Symmetries are broken by the conditions $m_X^2 > 0$ and $m_Y^2 > 0$ with $k < 0$, which corresponds to the case where instanton effects trigger both the chiral symmetry breaking and the $U_A(1)$ symmetry breaking. The fourth one is a new pattern which can be related to the study on the chiral symmetry breaking by instantons \cite{Diakonov:1995ea}.

To analyze the $0^{-+}$ sector, we use the nonlinear realization formula. By substituting the nonlinear parameterization of Eq. (\ref{goldstone}) into the Lagragian of Eq. (\ref{totl}), we obtain the kinematic terms of $\pi^0$, $\eta^0$, $\eta^8$, and $a$ fields, which determine their masses and how these pseudoscalar mix with eath other. Up to quadratic terms, the Lagrangian reads as
\bea
\cal{L} &= &\frac{1}{2} \partial \pi^0 \cdot \partial \pi^0 + \frac{1}{2} \partial \eta_q \cdot \partial \eta_q + \frac{1}{2} \partial \eta_s \cdot \partial \eta_s + \frac{1}{2} \partial a \cdot \partial a \\ \nonumber  &-& \frac{B_q}{2 v_q} (\pi^0)^2 - \frac{B_q}{2 v_q} (\eta_q)^2 - \frac{B_s}{2 v_s} (\eta_s)^2 - \frac{3 \, D}{2 v_Y} (a)^2 - \frac{1}{2} k v_Y v_q^2 v_s \left ( \frac{a}{v_Y} +\sqrt{\frac{2}{3}} \frac{\eta_q}{v_q} + \sqrt{\frac{1}{3}} \frac{\eta_s}{v_s} \right )^2 \,,
\eea
where we replace $\eta^0$ and $\eta^8$ by the combination of $\eta_q$ and $\eta_s$ 
\bea
\frac{\eta^0}{v_X}&=& \sqrt{\frac{2}{3} } \frac{ \eta_q}{v_q} + \sqrt{\frac{1}{3}} \frac{\eta_s}{v_s}\,, \\
\frac{\eta^8}{v_X} &=&  \sqrt{\frac{1}{3} } \frac{\eta_q}{v_q} -\sqrt{ \frac{2}{3}} \frac{\eta_s}{v_s}\,.
\eea
Obviously, neither $\pi^0$ nor $\eta^8$ mix with $a$ and $\eta^0$. Only the last term, originating from the instanton effects, leads to a mix between $\eta^0$ and the $0^{-+}$ glueball $a$. Such a mixing term splits the mass of  $\eta^0$ from that of the rest of pseudo-Goldstone bosons and provides a resolution to the $U_A(1)$ problem of QCD.

From the Lagrangian, it is straightforward to read out the mass matrix as
\begin{eqnarray}
\label{massm}
M^2_{gqs} = \left ( \begin{array}{ccc}
m_{gg}^2 + k v_q^2 \frac{ v_s} {v_Y}&  \sqrt{\frac{2}{3}} k v_q v_s  & \sqrt{ \frac{1}{3}} k  v_q^2  \\
\sqrt{\frac{2}{3}} k v_q v_s  &  m_{{\bar q} q}^2 + \frac{2}{3} k v_Y v_s  &  \frac{\sqrt{2}}{3} k v_Y v_q\\
 \sqrt{ \frac{1}{3}}  k v_q^2  &  \frac{\sqrt{2}}{3} k  v_Y v_q &
 m_{{\bar s}s}^2 + \frac{1}{3} k v_q^2 \frac{v_Y}{v_s}\\
\end{array} \right )\,.
\end{eqnarray} 
The $m_{gg}^2$ is equal to $3 \, D/v_Y$. In order to reduce the number of free parameters, we will assume that $B_0 = D_0$. We take the $m_{{\bar q} q}^2 =B/v_q = B_0 (m_u+m_d) = m_\pi^2$, and $m_{{\bar s} s}^2 = 2 M_k^2 - M_\pi^2$.

It is instructive to compare the mass matrix determined by the effective field theory method with that in the FKS formalism \cite{FKS}. Here we only consider the leading order appromixation. The terms in the first $2\times2$ components of the mass matrix can be compared directly to the FKS formalism ( We provide some crucial steps to derive the correspondence in Appendix ). We can have the following one-to-one correspondence:
\bea
m_{\q\q}^2 & =  & \frac{\sqrt2}{v_{\q}} \,
      \langle 0 | m_{\u} \, \bar{\u} \, i \gamma_5 \, {\u} +
                  m_{\d} \, \bar{\d} \, i \gamma_5 \, {\d} | \eta_{\q}
                  \rangle; \\
m_{\s\s}^2 & = & \frac{1}{v_{\s}} \,
      \langle 0 | m_{\s} \, \bar{\s} \, i \gamma_5 \, {\s} | \eta_{\s}
                  \rangle;\\
 \frac{\sqrt{2}}{3} k v_Y v_q &=&  \frac{\sqrt2}{v_{\q}}\,\langle 0|\frac{\alpha_s}{4\pi}\,G\tilde G|\eta_{\s} \rangle = \frac{1}{v_{\s}}\,\langle 0|\frac{\alpha_s}{4\pi}\,G\tilde G|\eta_{\q} \rangle; \\
\frac{2}{3} k  v_Y v_s &=& \frac{\sqrt2}{v_{\q}}\,
\langle 0|\frac{\alpha_s}{4\pi}\,G\tilde G|\eta_{\q}
  \rangle;\\
\frac{1}{3} k v_q^2 \frac{v_Y}{v_s}  & =&\frac{1}{v_{\s}}\,\langle 0|\frac{\alpha_s}{4\pi}\,G\tilde G|\eta_{\s}
  \rangle .
\eea
It is straightforward to have the following relations when FKS formalism is extended to include pseudoscalar glue ball
\bea
m_{gg}^2 + k  v_q^2 \frac{v_s}{v_Y} &=& \frac{\sqrt{3} }{v_{Y}} \,
      \langle 0 | \frac{\alpha_s}{4\pi}\,G\tilde G | a
                  \rangle;\\
\sqrt{\frac{2}{3}} k v_q v_s& =&   \frac{\sqrt{2}}{v_{q}} \,
      \langle 0 | \partial_\mu J^\mu_{5q}  | a
                  \rangle = \frac{\sqrt{3} }{v_{Y}} \,
      \langle 0 | \frac{\alpha_s}{4\pi}\,G\tilde G | \eta_q
                  \rangle;\\                 
\sqrt{\frac{1}{3}} k v_q^2 & =&   \frac{1}{v_{s}} \,
      \langle 0 | \partial_\mu J^\mu_{5s} | a
                  \rangle = \frac{\sqrt{3} }{v_{Y}} \,
      \langle 0 | \frac{\alpha_s}{4\pi}\,G\tilde G | \eta_s
                  \rangle;
\eea

Then we can have the following solutions in the leading approximation
\bea
m_{gg}^2 & =& 0 \,,\\
\langle 0 | \frac{\alpha_s}{4\pi}\,G\tilde G | \eta_q
                  \rangle & =& \frac{\sqrt{2} }{3} k v_Y v_q v_s\,, \label{vac2etaq}\\
\langle 0 | \frac{\alpha_s}{4\pi}\,G\tilde G | \eta_s
                  \rangle & =& \frac{ 1 }{3} k v_Y v_q^2 \,,\label{vac2etas}\\
\langle 0 | \frac{\alpha_s}{4\pi}\,G\tilde G | a
                  \rangle & =&  \langle 0 | \partial_\mu J^\mu_{5q} | a
                  \rangle =  \langle 0 | \partial_\mu J^\mu_{5s} | a
                  \rangle = \sqrt{ \frac{1}{3} } k v_q^2 v_s \,.\label{vac2a}
\eea

Thanks to the relation given in Eq. (\ref{vac2a}), it is possible to extract the value of $k$ from the lattice calculation. Due to the relation between the effective field theory and the FKS formula, it is also possible to extract the value of $k$ from the pseudoscalar meson spectra. In the following section, we will discuss the computation of $k$ in QCD by considering the instanton effects and its uncertainties.

\section{The $k$ parameter and the effective gluon mass $m_g^{eff}$}
\setcounter{equation}{0}
\renewcommand{\theequation}{\arabic{section}.\arabic{equation}}%

The generating functional induced by one instanton can be deduced 
from QCD, as shown by 't Hooft \cite{'tHooft:1976fv}, which reads
\bea
\int_x k_t e^\theta \det_f^{N_f} \left ( \bar{ \psi_L^f}(x) \psi_R^f(x) \right )\,,
\eea
where $\psi$ denotes the quark fields, $k_t$ is a constant, $x$ is the position of instanton, the $f$ denotes the index of flavor space, and $\theta$ is the vacuum angle. Following the de-clustering assumption and summing over all the contributions of both instantons and anti-instantons, we can arrive at the effective interaction which reads as
\bea
\exp\left \{ \int_x k_t e^\theta \det_f^{N_f} \left ( \bar{ \psi_L^f}(x) \psi_R^f(x) \right )  + h.c.   \right \}\,.
\eea
To relate this effective interaction with that of the meson fields, we need redefine the following variable
\bea
\phi(x) =  \frac{ \bar{ \psi_L^f}(x) \psi_R^f(x) } {\Lambda^2_{match}}\,,
\eea
where $\Lambda_{match}$ is the matching scale between the fundamental theory and the effective theory, which is introduced from the dimensional analysis. Then we can express the effective interaction of mesons induced from instantons as
\bea
\exp\left \{ \int_x k_t \Lambda^{2 N_f}_{match} e^\theta \det_f^{N_f} \left ( \phi(x) \right )  + h.c.   \right \}\,.
\eea
In order to formulate the solution of the $U_A(1)$ problem within the framework of spontaneous symmetry breaking, it is instructive to introduce a field $Y$ to realize the continuous symmetry \cite{'tHooft:1986nc}. Therefore the effective interaction arrives at the form
\bea
\exp\left \{ \int_x \frac{k}{6} Y \det_f^{N_f} \left ( \phi(x) \right )  + h.c.   \right \}\,,
\eea
which is given in Eq. (\ref{model}), where the effective coupling $k$ is expressed by $k_t$ as
\bea
k & = & \frac{ 6 k_t \Lambda^{2 N_f}_{match} } {v_Y}\,.
\eea
Implicitly, this $Y$ field is related with the instanton configurations of QCD. Compared with other forms of effective interaction proposed in literatures, the advantage of this interaction is that the anomalous Ward identities can be easily satisfied.

Now we address the strength of the effective coupling $k_t$. $k_t$ can be obtained from the one-loop calculation. For the $SU(N_C)$ group, the contribution of gluons can be formulated as
 \cite{Bernard:1979qt} \bea
k_t^G &=& \rho^{-5} \frac{4}{\pi^2} \frac{ e^{-\alpha(1) - 2 (N_c - 2) \alpha(\frac{1}{2}) }}{(N_c-1)!(N_c-2)!}  \left ( \frac{4 \pi^2}{g(\mu)^2} \right )^{2 N_c} e^{ - \frac{8 \pi^2}{g^2} +  \frac{11}{3} N_c ln(\mu \rho) }\,,
\eea
where $N_c$ is the number of color. Obviously, in the large $N_c$ limit, $k_t$ vanishes. The contribution of $N_f$ chiral quarks can be formulated as \cite{'tHooft:1976fv}
\bea
k_t^q &=&2^{3 N_f} \pi^{2 N_f} \rho^{3 N_f} e^{-\frac{2}{3} N_f ln(\mu \rho) - 2 N_f \alpha(\frac{1}{2}) } \,.
\eea
Combining the contributions of gluons and quarks (the integration of the product of $k_t^G$ and $k_t^q$ over instanton size $\rho$), we arrive at
\bea
k_t &=& C_k \int_\rho \rho^{3 N_f - 5}  \left ( \frac{4 \pi^2}{g(\mu)^2} \right )^{2 N_c} e^{ -\frac{8 \pi^2}{g(\rho)^2} }\,,\label{kteq}\\
C_k &=& \frac{4}{\pi^2} 2^{3 N_f} \pi^{2 N_f} \frac{ e^{-\alpha(1) - 2 (N_c - 2) \alpha(\frac{1}{2}) + 2 N_f \alpha(\frac{1}{2}) } }{(N_c-1)!(N_c-2)!}\,.
\eea
To obtain this expression, we have utilized the renormalization group equation of gauge coupling
\bea
\frac{8 \pi^2}{g(1/\rho)^2 } = \frac{8 \pi^2}{ g(\mu)^2} - \left[ \frac{11}{3} N_c - \frac{2}{3} N_f \right] ln(\mu \rho)\,.
\eea
While the constants $\alpha(1)$ and $\alpha(\frac{1}{2})$ are dependent on the regularization scheme and are given in \cite{'tHooft:1986nc}. According to the study of Geshkenbein and Ioffe in \cite{Geshkenbein:1979vb}, it is better to use the results of dimensional regularization. Then these two constants are given as
\bea
\alpha(1) & = & 8 R + \frac{1}{3} ln 2 - \frac{16}{9}\,, \\
\alpha(\frac{1}{2}) &=& 2 R - \frac{1}{6} ln2- \frac{17}{72}\,,\\
R&=& \frac{1}{12} \left [ log(2 \pi) + \gamma \right]  + \frac{1}{2 \pi^2} \sum_2^{\infty} \frac{log~s}{s^2} \\ \nonumber
&=& 0.248754\,.
\eea

Below we address the uncertainty of $k_t$. There are a couple of theoretical uncertainties in determining the value of $k_t$: 1) the first uncertainty is from the regularization scheme. For instance, the value of $k_t$ in dimensional regularization scheme can be one thousand times larger than its value in Pauli-Vilas regularization scheme. 2) The second uncertainty is from the infrared cutoff in the size of instantons. 

In Table (\ref{tablek}), we tabulate the dependence of $k_t$ on the infrared cutoff $\rho_{IR}$ of the size of instantons. In order to explore the dependence of $k_t$ on the energy scale, we divide the integration of Eq. (\ref{kteq}) into three interval integration: 
\bea
k_t^1 &=& {\int}_{0}^{\frac{1}{m_b}}  f_K(N_c, N_f) d\rho \,, \\
k_t^2 &=& {\int}_{\frac{1}{m_c}}^{\frac{1}{m_b}} f_K(N_c, N_f)  d\rho \,,\\
k_t^3 &=& {\int}_{\frac{1}{m_c}}^{\rho_{IR}} f_K(N_c, N_f) d\rho\,,
\eea
where $m_b$ and $m_c$ are the masses of $b$ and $c$ quarks. The final value of $k_t$ is the sum of these three parts. The integrand function $ f_K(N_c, N_f)$ can be read out from Eq. (\ref{kteq}). In these three integration intervals, the effective active degrees of quark flavors, $N_f$, are different. 

There are several comments on the Table (\ref{tablek}) in order. 1) One fact is that the dominant contribution to $k_t$ is from the infrared regions as demonstrated by the value of $k_t^1$ and $k_t^2$. 
2) The value of $k_t$ is very sensitive to the change of $\rho_{IR}$, for instance, when $\rho_{IR}$ changes from $1$ to $2$, the value of $k_t^3$ increases by $10^3$ times, as shown in the 4th and 5th columns. 3) Values of $k_t$ in the dimensional regularization scheme are larger than those in the Pauli-Villas regularization scheme by a factor $e^{8.8}$. 

\begin{table}[th]
\begin{center}%
\begin{tabular}
{|c|c|c|c|c|c|c|}\hline &  &  &  & & & \\
Contri to $k_t$& $k_t^1$& $k_t^2$ &
 $k_t^3$ &$k_t^3$
&$k_t$&$k_t$ \\
&&&$\rho_{IR} =1$ &$ \rho_{IR} =2 $&$\rho_{IR}=1$&$\rho_{IR}=2$\\
\hline
\hline
&  &  &  & & & \\
 Dim sheme& $1.30\times {10}^{-4}$ & $1.07\times 10^{3}$ & $7.45\times10^1 $ & $1.11 \times10^6$&$1.15\times10^3$&$1.11\times 10^6$ \\
&  &  &  &  & &\\\hline 
&  &  &  &  & & \\
P-V scheme  & $1.97\times 10^{-8}$ & $1.61\times10^{-1}$ &
$1.12\times10^{-2}$ &$1.68 \times 10^2$&$1.73\times10^{-1}$&$1.68\times10^2$\\
&  &  &  &  & & \\\hline
\end{tabular}
\end{center}
\caption{ The values of $k_t$ in different $\rho_{IR}$s and regularization schemes are shown. The Dim
and P-V schemes denote for the dimensional regularization scheme and the Pauli-Vilas regularization
scheme, respectively. The $\rho_{IR}$ has the unit as GeV$^{-1}$.}%
\label{tablek}%
\end{table}

The value of $k$ further suffers an uncertainty from the determination of $\Lambda^{2 N_f}_{match}$ and $v_Y$. Below we will assume that $\Lambda_{match} = 1$ GeV.

Table \ref{tablek} explicitly shows that it is difficult to determine the value of $k_t$ and $k$ from pure theoretical side due to our ignorance on the value of $\rho_{IR}$. If we believe that the dimensinal regularization is the correct scheme and to have $k  \sim 60 $, it seems that the $\rho_{IR}$ should be around $0.5$ GeV$^{-1}$ or so, which corresponds to $\Lambda_{IR} \approx 2$ GeV. Here $\Lambda_{IR}$ is the matching scale of QCD and the effective chiral Lagrangian given in Eq. (\ref{totl}). While in the Pauli-Villas regularization, the predicted $k$ is too small to solve the $U_A(1)$ problem.

From now on, we address the effective gluon mass $m_g^{eff}$.
It is well known that in the perturbation and large momentum region, the mass of gluon vanishes. In the infrared region and large distance region, the effective mass of gluon might be non-vanishing. Currently there are two methods to extract the effective mass of gluon: one is from the Dyson-Schwinger equations as demonstrated in references \cite{'A.C.Aguilar and J.Papavassilion:2008nc,Aguilar:2008xm}, the other is from the lattice calculation (as demonstrated in references \cite{gupta87,mandula99,I.L.Bogolubsky:2009zz}).

From the method of the Dyson-Schwinger equations, as shown in references \cite{'A.C.Aguilar and J.Papavassilion:2008nc,Aguilar:2008xm}, the gluon propagator can be computed and the effective mass of gluon can be extracted. Depending on the ansatz of the three-gluon vertex, two types of solutions for the effective mass of gluon can be derived. The first type of solutions can be parameterized as
\bea
\left ( m_g^{eff} \right )^2(q^2)&=& {m_0}^2[\ln(\frac{q^2+\rho
{m_0}^2}{\Lambda^2})/\ln(\frac{\rho
{m_0}^2}{\Lambda^2})]^{-1-\gamma_1}\,,
\eea
then the effective squared mass $\left ( m_g^{eff} \right )^2(q^2)$ has a logarithmic running behavior, while $\gamma_1=\frac{6}{5}(1+c_2-c_1)$. The second type of solutions reads as
\bea 
\left ( m_g^{eff} \right ) ^2 (q^2)&=& \frac{{m_0}^4}{q^2+{m_0}^2}[\ln(\frac{q^2+\rho
{m_0}^2}{\Lambda^2})/\ln(\frac{\rho
{m_0}^2}{\Lambda^2})]^{\gamma_2-1}\,,
\eea
then the effective squared mass $\left ( m_g^{eff} \right )^2(q^2)$ has a power law running behavior, while $\gamma_2=\frac{4}{5}+\frac{6c_1}{5}$. The $\Lambda$ is the mass scale of QCD, which is taken as $\Lambda=0.30$ GeV.

\begin{figure}[t]
\centerline{ \epsfxsize=6.0 cm \epsfysize=5.5cm
\epsfbox{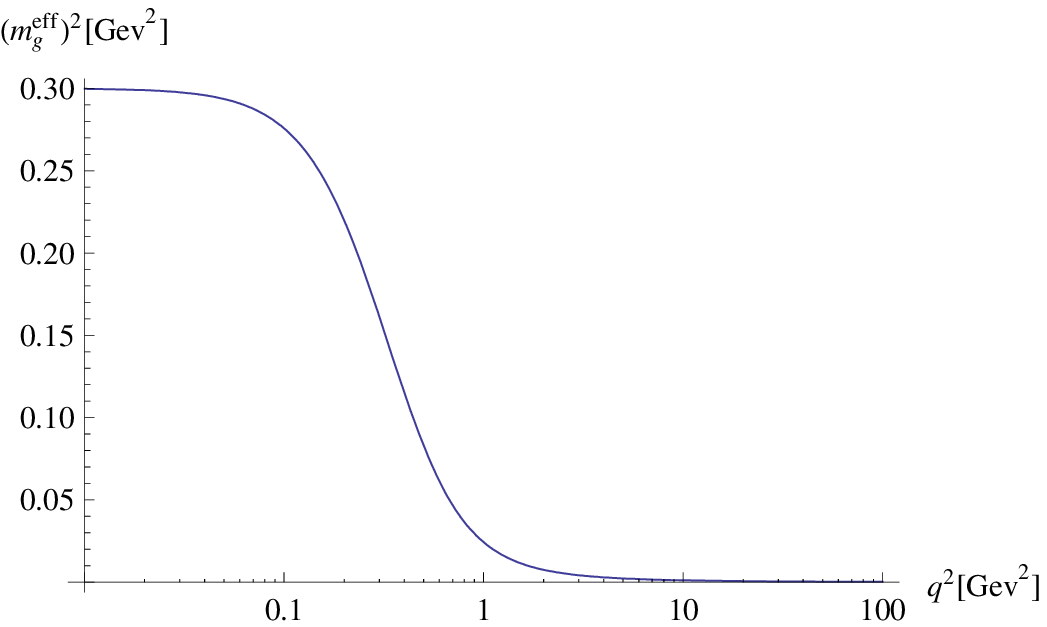} \hspace*{0.1cm} \epsfxsize=6.0 cm
\epsfysize=5.5cm \epsfbox{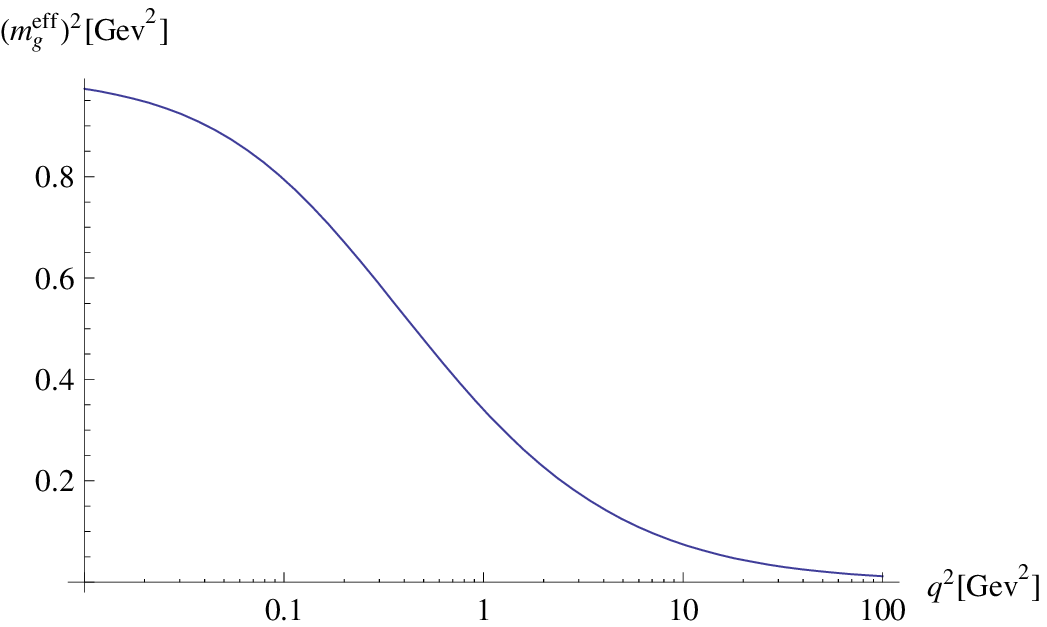} }\vskip -0.05cm \hskip 0.4
cm \textbf{( a ) } \hskip 4.8 cm \textbf{( b )} 
\caption{\textit{The dependence of the effective gluon mass $m^2$ on the external momentum $q^2$ is shown. Plot (a) depicts the case where $m^2$ has the power law running behavior, and
plot (b) depicts the case where $m^2$ has the logarithmic running behavior. 
}} 
\label{fig-bd}
\end{figure}

We show these two types of solutions in Fig. (\ref{fig-bd}). In Fig. (\ref{fig-bd}a), the parameters are taken as $m_0^2=0.3$ GeV$^2$, $\rho=1.007$, $c_1=0.15$, and $c_2=-0.9635$.
In Fig. (\ref{fig-bd}b), the parameters are taken as $m_0^2=0.5$ GeV$^2$, $\rho=1.046$, $c_1=1.1$, and $c_2=-1.121$. 

From Fig. (\ref{fig-bd}) and the analysis given in reference \cite{'A.C.Aguilar and J.Papavassilion:2008nc}, we can read out that the effective mass of gluon can expand in a range $0.5$ GeV $<m_g^{eff} < 1.2$ GeV when $q^2 \to 0$.

The lattice method can also extract the effective mass of gluon. For instance, the references \cite{Attilio Cucchieri:2007zz,I.L.Bogolubsky:2009zz} present results where the gluon mass in $SU(3)$ case is $0.3 $ GeV $\sim 0.5$ GeV or so. The results are claimed to be consistent with the so-called decoupling solutions found for Dyson-Schwinger and functional renormalization group equations. Meanwhile, as shown in \cite{A. Sternbeck:2007zz}, it was found that there is very little $N_c$-dependence in the gluon's propagator over the whole momentum range, so is the effective mass of gluon. 

Combining the results given by these two computational methods, in this paper, we will assume that the effective mass of gluon is in the range $0.3$ GeV $< m_g^{eff} < 1.2$ GeV.

\section{Numerical Analysis}
\setcounter{equation}{0}
\renewcommand{\theequation}{\arabic{section}.\arabic{equation}}%
To do numerical analysis, we will take the following inputs: 
\bea
\label{inputs}
v_q=0.092 {\rm GeV}\,\,,\,\,\,\,v_s = \sqrt{2} v_q\,\,,\,\,\,\,m_u=0.003 {\rm GeV}\,\,,\,\,\,\,m_d=0.007 {\rm GeV}\,\,,\\\nonumber
m_s=0.15 {\rm GeV}\,\,,\,\,\,\,m_{\pi}=0.1395 {\rm GeV}\,\,,\,\,\,\,m_{K}=0.493 {\rm GeV}\,\,.
\eea

Below we concentrate on the analysis of the allowed parameter space of $m_g^{eff}$, $k$, and $k \,\, v_Y$. First we study how the $0^{++}$ can constrain the parameter space. To scan the parameter space, we allow the effective gluon mass to vary in the range $0.3$ GeV $< m_g^{eff} < 1.2$ GeV, the vacuum expectation value $v_Y$ to vary in the range $ 0.001$ GeV $< v_Y < 1$ GeV for the first scenario and $0.001$ GeV $< v_Y < 1.5$ GeV for the second scenario, and the mixing angle to vary in the range $-\frac{\pi}{2}< \alpha < \frac{\pi}{2}$.

\begin{figure}[t]
\centerline{
\epsfxsize=6.5 cm \epsfysize=6.0cm \epsfbox{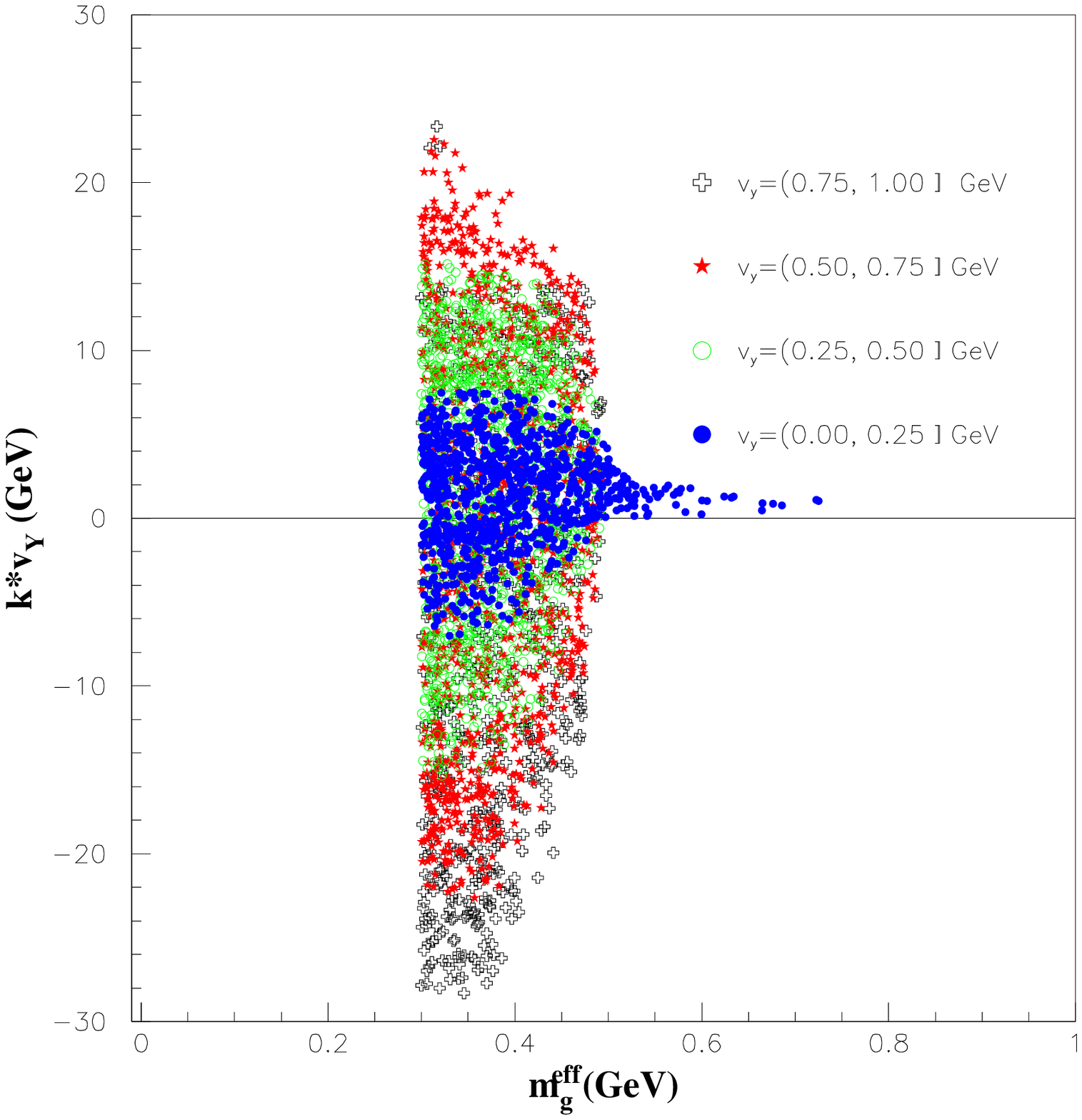}
\hspace*{0.4cm}
\epsfxsize=6.5 cm \epsfysize=6.0cm \epsfbox{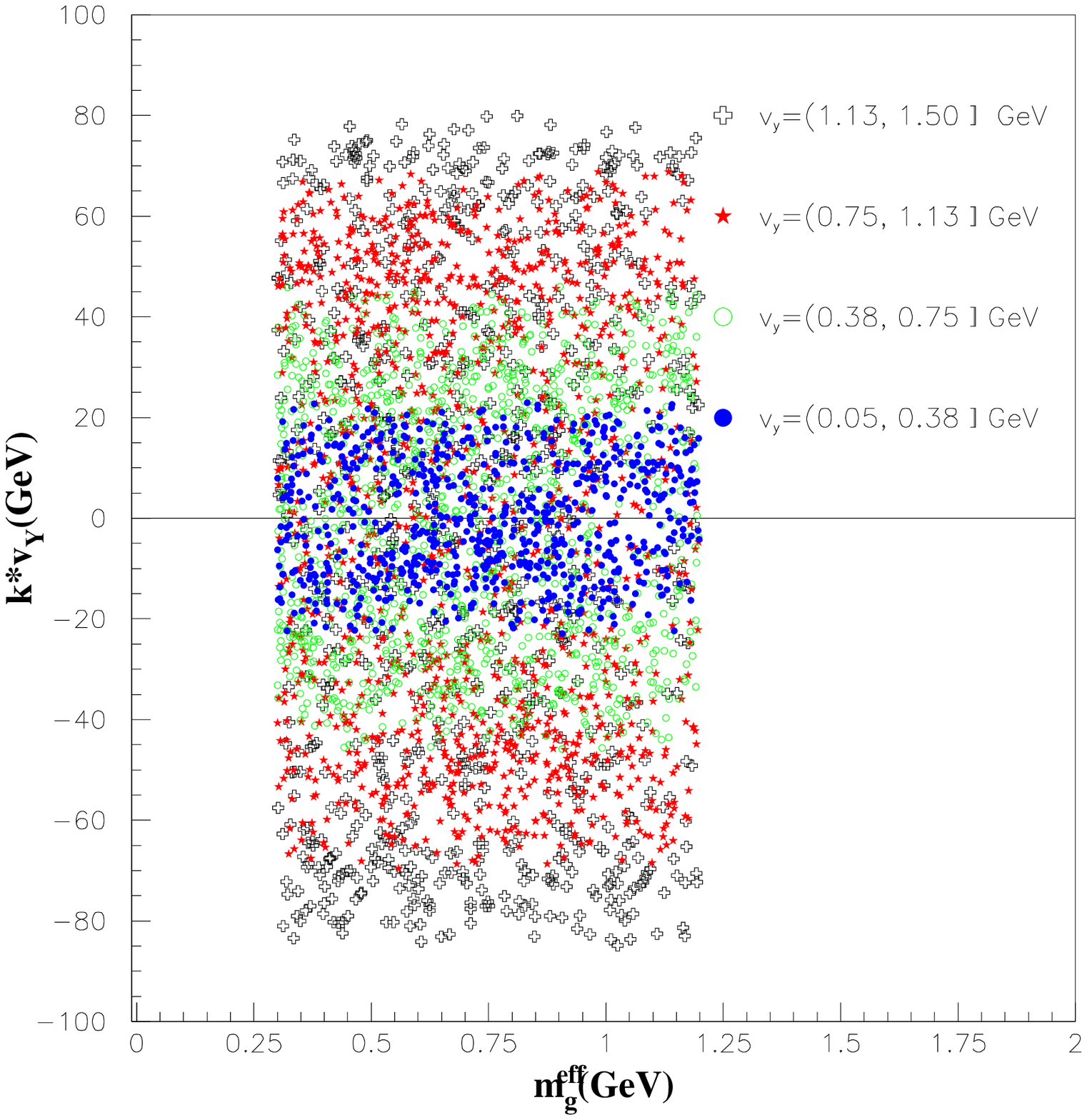}
}
\vskip -0.05cm \hskip 0.0 cm \textbf{( a ) } \hskip 6.6 cm \textbf{( b )}
\centerline{
\epsfxsize=6.5 cm \epsfysize=6.0cm \epsfbox{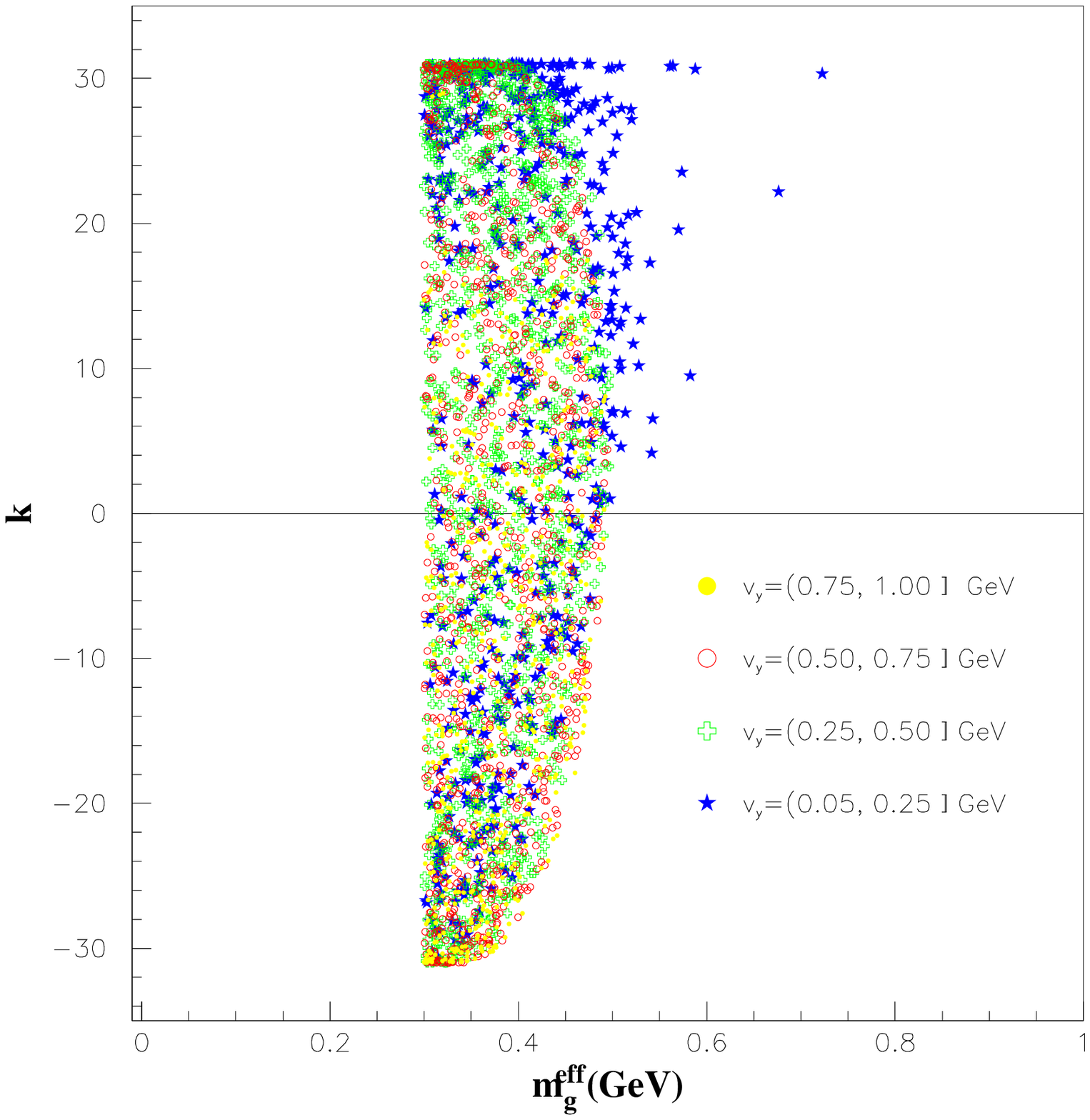}
\hspace*{0.2cm}
\epsfxsize=6.5 cm \epsfysize=6.0cm \epsfbox{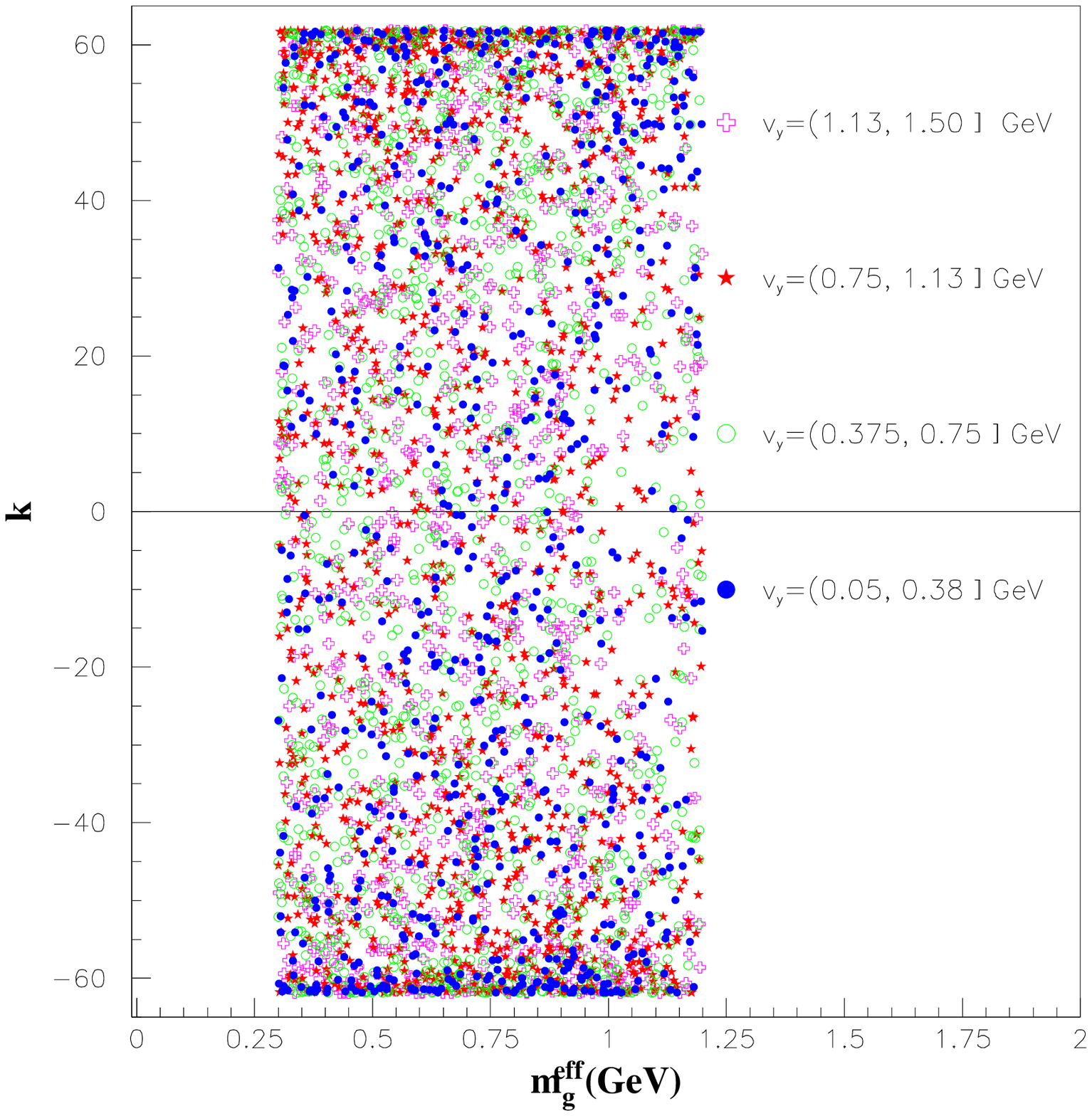}
}
\vskip -0.05cm \hskip 0.0 cm \textbf{( c ) } \hskip 6.6 cm \textbf{( d )}
\caption{\textit{The parameter regions from the constraints of vacuum conditions and the $0^{++}$ sector are shown. Plots (a) and (c) are devoted for the scenario one, and plots (b) and (d) for the scenario two.}}
\label{fig-vac}
\end{figure}

Fig. (\ref{fig-vac}) shows the parameter space allowed by the vacuum constraints, untarity bounds, and the $0^{++}$ inputs. Comparing \ref{fig-vac}(a) and \ref{fig-vac}(b), we observe that the allowed region for $m_g^{eff}$ and $k\, v_Y$ is larger in the second scenario than that in  the first scenario. It is similar for $k$, as demonstrated by  \ref{fig-vac}(c) and \ref{fig-vac}(d). One interesting observation is that the value of $k$ can be either positive or negative, which can realize the fourth pattern of the chiral symmetry and $U_A(1)$ symmetry breaking.

Now we study the parameter space allowed by the $0^{-+}$ sector and treat $k$, $v_Y$, and $m_g^{eff}$ as parameters with no bounds. Substituting the inputs given in Eq. (\ref{inputs}) into the mass matrix of pseudoscalar, we can arrive at the following mass matrix which describes the mixing of pure $0^{-+}$ glueball state, $q{\bar q}$ state, and $s {\bar s}$ state:
\begin{eqnarray}
\label{massmgqs}
M^2_{gqs} = \left ( \begin{array}{ccc}
1.0 \times 10^{-3} \frac{k}{v_Y} + 1.95 \, m_g^{eff} &   9.8\times 10^{-3} \, k & 4.9 \times 10^{-3} \, k  \\
9.8\times 10^{-3} k &  1.95 \times 10^{-2} + 8.67\times 10^{-2} \, k \, v_Y  & 4.34 \times 10^{-2} \, k \, v_Y \\
 4.89 \times 10^{-3} \, k  & 4.34 \times 10^{-2} \, k \, v_Y &
 0.47 + 2.17 \times 10^{-2} \, k \, v_Y\\
\end{array} \right )\,.
\end{eqnarray} 

In order to make the solution of the $U_A(1)$ problem explicitly, it is also useful to rewrite the mass matrix in the basis of $a$, $\eta_0$, and $\eta_8$. The matrix reads as
\begin{eqnarray}
\label{massmg08}
M^2_{g08} = \left ( \begin{array}{ccc}
1.0 \times 10^{-3} \frac{k}{v_Y} + 1.95 \, m_g^{eff} &   1.0\times 10^{-2} \, k & 1.6 \times 10^{-3} \, k  \\
1.0\times 10^{-2} k &  0.168 + 0.1 \, k \, v_Y  & - 0.21 + 1.6\times 10^{-2} \, k \, v_Y \\
 1.6 \times 10^{-3} \, k  & -0.21 + 1.6 \times 10^{-2} \, k \, v_Y &
 0.32 + 2.5 \times 10^{-3} \, k \, v_Y\\
\end{array} \right )\,.
\end{eqnarray} 
This numerical mass matrix reveals several transparant but interesting facts on the solution of the $U_A(1)$ problem, which will be listed in order below.
1) Comparing the $22$ and $33$ element, we can know that if $k=0$, then $m_{\eta^0}$ is always smaller than $m_{\eta^8}$. 
2) Generally speaking, if $k<<10^3$, the mixing of pure glueball $a$ and $\eta^8$ is negligibly small, which indeed supports the two angle scheme to describe the mixing among pure glueball state $a$, $\eta_{q{\bar q}}$ and $\eta_{s{\bar s}} $, as demonstrated by $12$ and $13$ elements.
3) In order to solve the $U_A(1)$ problem, the product of $k$ and $v_Y$ must be large enough (say $8 \sim 13$) to lift the $m_{\eta^0}$ and realize the scenario with $m_{\eta^0} > m_{\eta^8}$, as indicated by the $22$ element. While due to the suppression factor $2.5 \times 10^{-3}$ in the $33$ element, the mass of $\eta^8$ is not sensitive to the change of $k \,\, v_Y$.
4) From the $23$ element, we can read that there is a minimum mixing scenario for $\eta^0$ and $\eta^8$, which correspond to $k\, v_Y \approx 13$.
5) If $m_g^{eff}$ is very large, the glueball state might decouple from the solution of $U_A(1)$ problem.

In order to examine how the parameter $k$ controls the spectra and mixing, we select the following two bench mark cases to study the dependence of mass spectra and mixing on parameters. These two bench mark cases assume that $\eta(1405)$ and $\eta(2190)$ are glueball candidates, respectively. The values of $v_Y$ and $m_g^{eff}$ are selected from the solutions by using the least $\chi^2$ method. Such solutions are possible since the mass matrix of pseudoscalar includes three free parameters while we can require that the eigenvalues of the mass matrix correspond to the masses of $\eta(548)$, $\eta^\prime(958)$, and the mass of the physics glueball candidate, which is equivalent to impose three conditions. Table \ref{table-bench} lists the values of these two bench make cases. 

\begin{table}[th]
\begin{center}%
\begin{tabular}
[c]{|c|c|c|c|}\hline
& & & \\
& $v_Y$ & $m_g^{eff}$& $k$  \\
& & & \\ \hline
& & & \\
Bench case 1 $\eta(1405)$ & $0.234$  & $0.824$ & $37.75$ \\
& & & \\ \hline
& & & \\
Bench case 2 $\eta(2190)$ & $0.133$  & $2.160$ & $58.72$ \\
& & & \\ \hline
\end{tabular}
\end{center}
\caption{ Two bench mark cases to show how the spectra of 
pseudoscalars and their fraction are affected by the parameters. The best fit parameters are determined by using the least $\chi^2$ method. We select only one solution for each bench case, though the solution for each case is not unique.}%
\label{table-bench}%
\end{table}

\begin{figure}[t]
\centerline{
\epsfxsize=6.5 cm \epsfysize=4.5cm \epsfbox{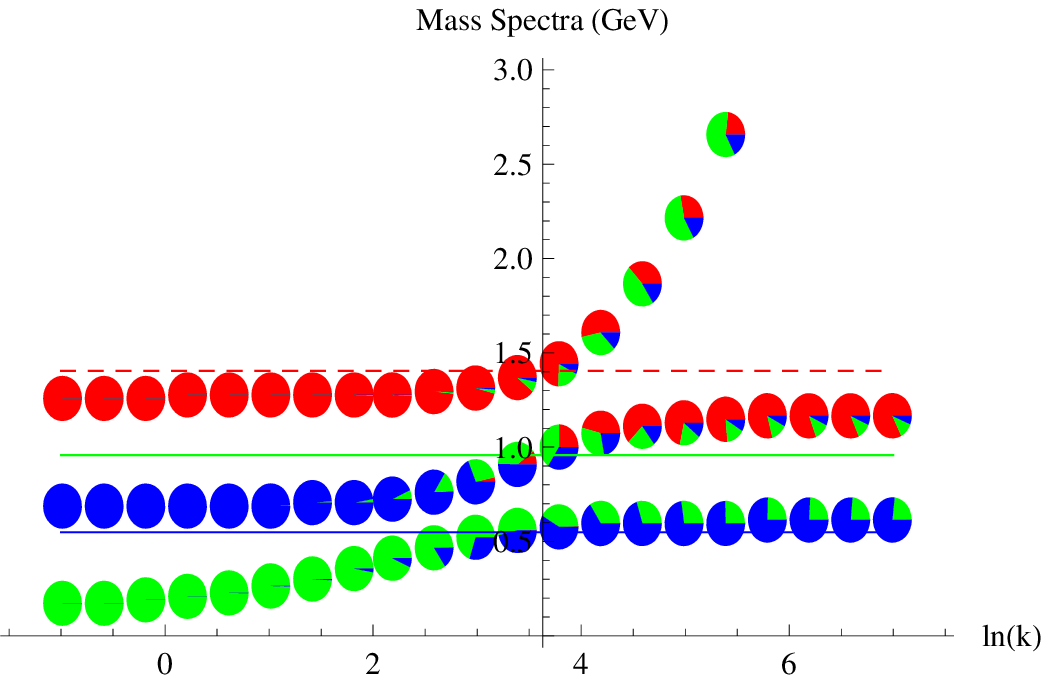}
\hspace*{0.2cm}
\epsfxsize=6.5 cm \epsfysize=4.5cm \epsfbox{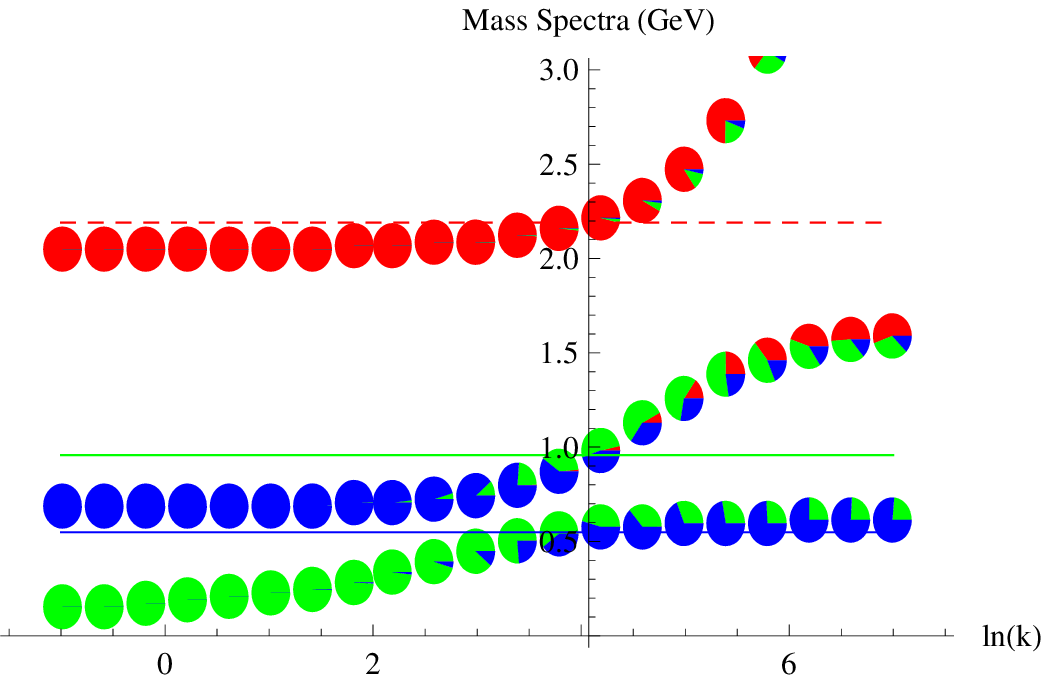}
}
\vskip -0.05cm \hskip 0.0 cm \textbf{( a ) } \hskip 6.6 cm \textbf{( b )}
\vskip 0.1cm
\centerline{
\epsfxsize=6.5 cm \epsfysize=4.5cm \epsfbox{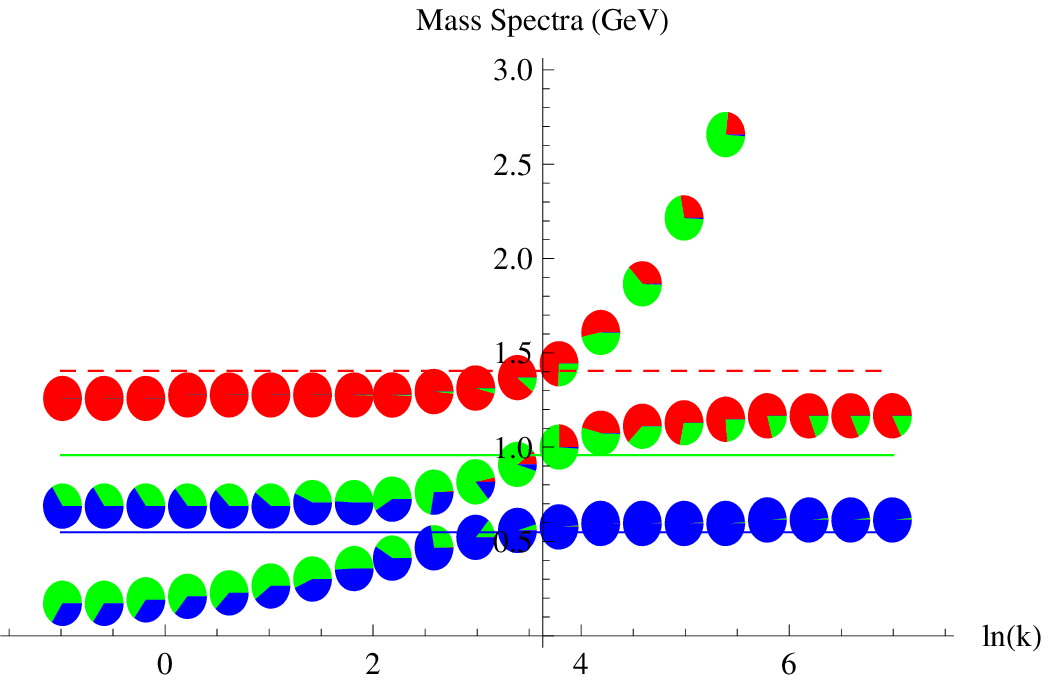}
\hspace*{0.2cm}
\epsfxsize=6.5 cm \epsfysize=4.5cm \epsfbox{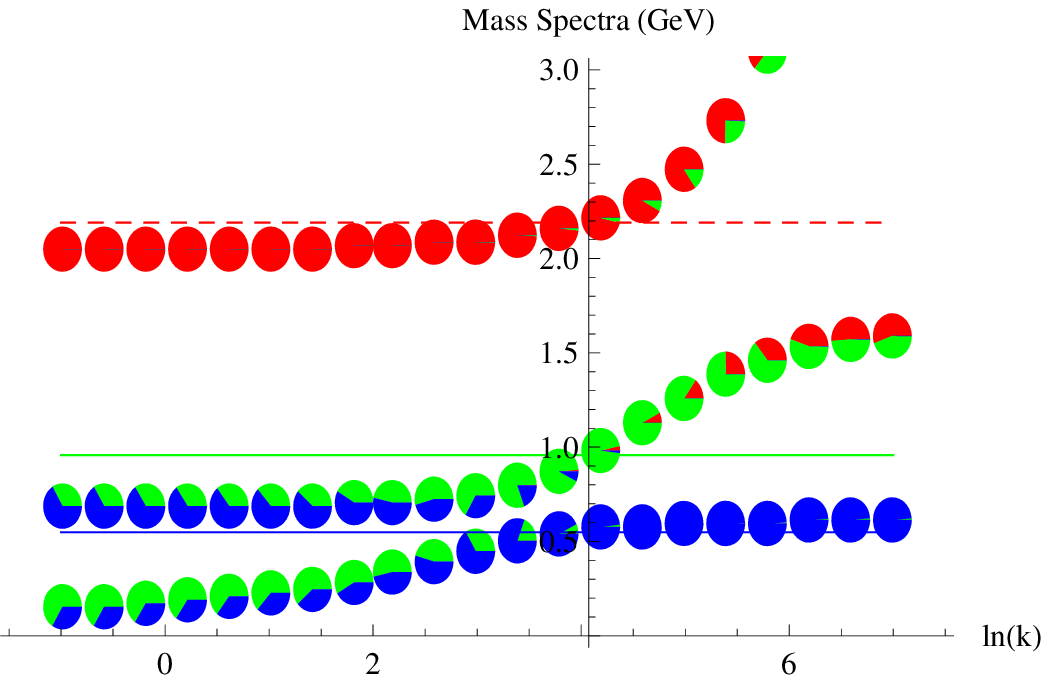}
}
\vskip -0.05cm \hskip 0.0 cm \textbf{( c ) } \hskip 6.6 cm \textbf{( d )}
\caption{\textit{The dependence of the spectra and their fractions on parameter $k$ is shown. The y axis is the mass of the spactra, and the x axis is the parameter $k$. Each pie in the plot denotes the component of the glueball, $q {\bar q}$ ( or $\eta^0$ ), and $s {\bar s}$ ( or $\eta^8$ ) states. The figures (a) and (b) are in the basis of $a$, $q{\bar q}$, and $s {\bar s}$. The figure (c) and (d) are in the basis of $a$, $\eta^0$, and $\eta^8$. The red part of each pie denotes the glueball component. For figures (a) and (b), the green part denotes the $q {\bar q}$ component, and the blue one denotes the $s {\bar s}$ component. For figures (c) and (d), the green part denotes the $\eta^0$ component, and the blue one denotes the $\eta^8$ component. The y axis is deliberately chosen to located in the best fitted $k$ for each plot. The red, blue and green lines represent the experimental values of $\eta(548)$, $\eta(958)$, and $\eta(1405)$ (or $\eta(2190)$).}}
\label{fig-9bm}
\end{figure}

In Fig. (\ref{fig-9bm}), we use a pie with three colors to denote the three components of each spectrum: red, blue, and green denote glue ball, $q {\bar q}$ (or $\eta^0$), and $s {\bar s}$ (or $\eta^8$), respectively. The fraction of these three components is determined by the eigenvector of each eigenstate. For instance, if the eigenstate of $\eta(548)$ is found to be $|\eta(548)\rangle = V_{\eta g}|gg \rangle + V_{\eta q}|q {\bar q}\rangle + V_{\eta s} |s{\bar s}\rangle$, where $V_{\eta i}, i=g, q, s$ is  defined as the component of eigenvector. These components must satisfy the normalization condition as $\sum_i V_{\eta i}^2 =1$. The fraction of each component in the pie is determined by $V_{\eta i}^2$.
Fig. (\ref{fig-9bm}a) and Fig. (\ref{fig-9bm}c) depict the bench case for  $m_\eta=1.405$ GeV, and Fig. (\ref{fig-9bm}b) and Fig. (\ref{fig-9bm}d) for $m_\eta=2.190$ GeV. 

From these plots, we can read out the fact that when $k$ is small (say  $k<15$ or so), the mixing is negligibly small and the mass eigenstates are almost pure states. Then the state with dominant $q {\bar q}$ (or $\eta^0$) component is lighter than that with dominant $s {\bar s}$ (or $\eta^8$). When $k$ increases (say $15<k<60$) , the mass of $q {\bar q}$ (or $\eta^0$) state increases quickly, while the mass of $s {\bar s}$ (or $\eta^8$) state is not sensitive to the change of $k$. When $k$ is large enough, the mixing between $q {\bar q}$ and $s {\bar s}$ becomes significant and can realize the masses of $\eta(548)$ (which is dominant $\eta^8$ component) and $\eta^\prime(958)$. If $k$ increases further (say $k>60$), the lightest state ($\eta^8$) is almost insensitive to the change of $k$, while the masses of the heaviest and second heaviest states become larger. It is also noticed that when $k>60$, the heaviest state can be either glueball dominant state or the $q {\bar q}$ dominant state. 

We also observe that in the solutions, the $\eta^\prime(958)$ contains more glueball component in the bench case one than in the bench case two. The bench case two demonstrates the decoupling limit of glueball state in the solution of the $U_A(1)$ problem. This fact indicates that if glueball candidate is light (say less than 1500 MeV), $\eta^\prime(958)$ might has considerable amount of glueball component. While if glueball candidate is heavy (say heavy than 2000 MeV as predicted from lattice computation), then $\eta^\prime(958)$ might has no significant amount of glueball component. However, such an amount of glueball component in $\eta^\prime(958)$ depends on the parameters of the solutions.

For the solutions, we observe one interesting fact: in order to solve the $U_A(1)$ problem, the heaviest state is always the glueball dominant state, and the second heaviest state is always $\eta^0$ dominant state. There is no solution for the case where $\eta^\prime$ is dominantly glueball component.

From these two bench cases, we know how the $U_A(1)$ problem is solved and the role played by the glueball candidates. Here and below, we extend to study the region of parameter for nine possible psedudo-scalar glueball candidates, as given in Table \ref{table1}.
\begin{figure}[t]
\centerline{
\epsfxsize=6.5 cm \epsfysize=6.0cm \epsfbox{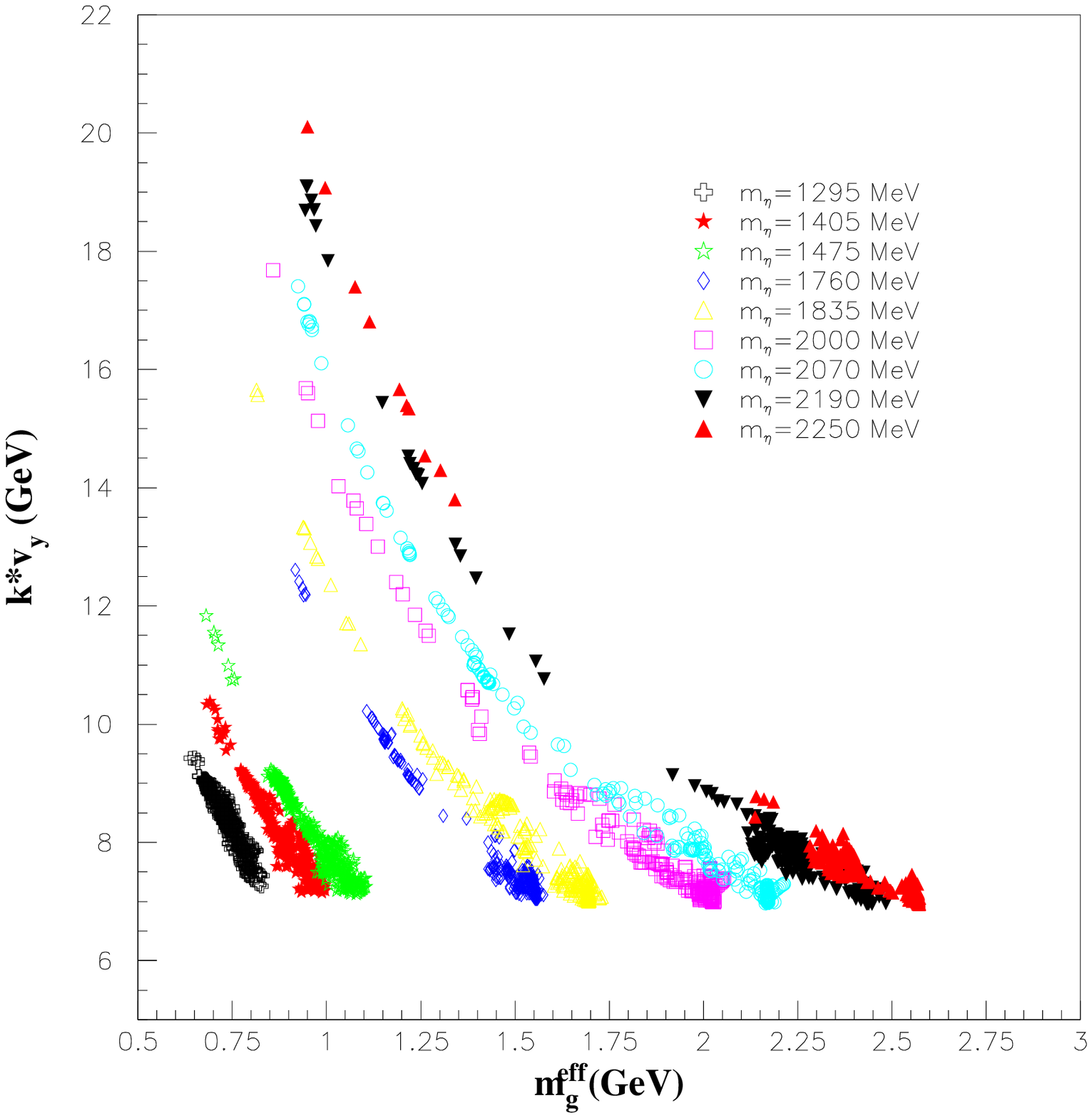}
\hspace*{0.2cm}
\epsfxsize=6.5 cm \epsfysize=6.0cm \epsfbox{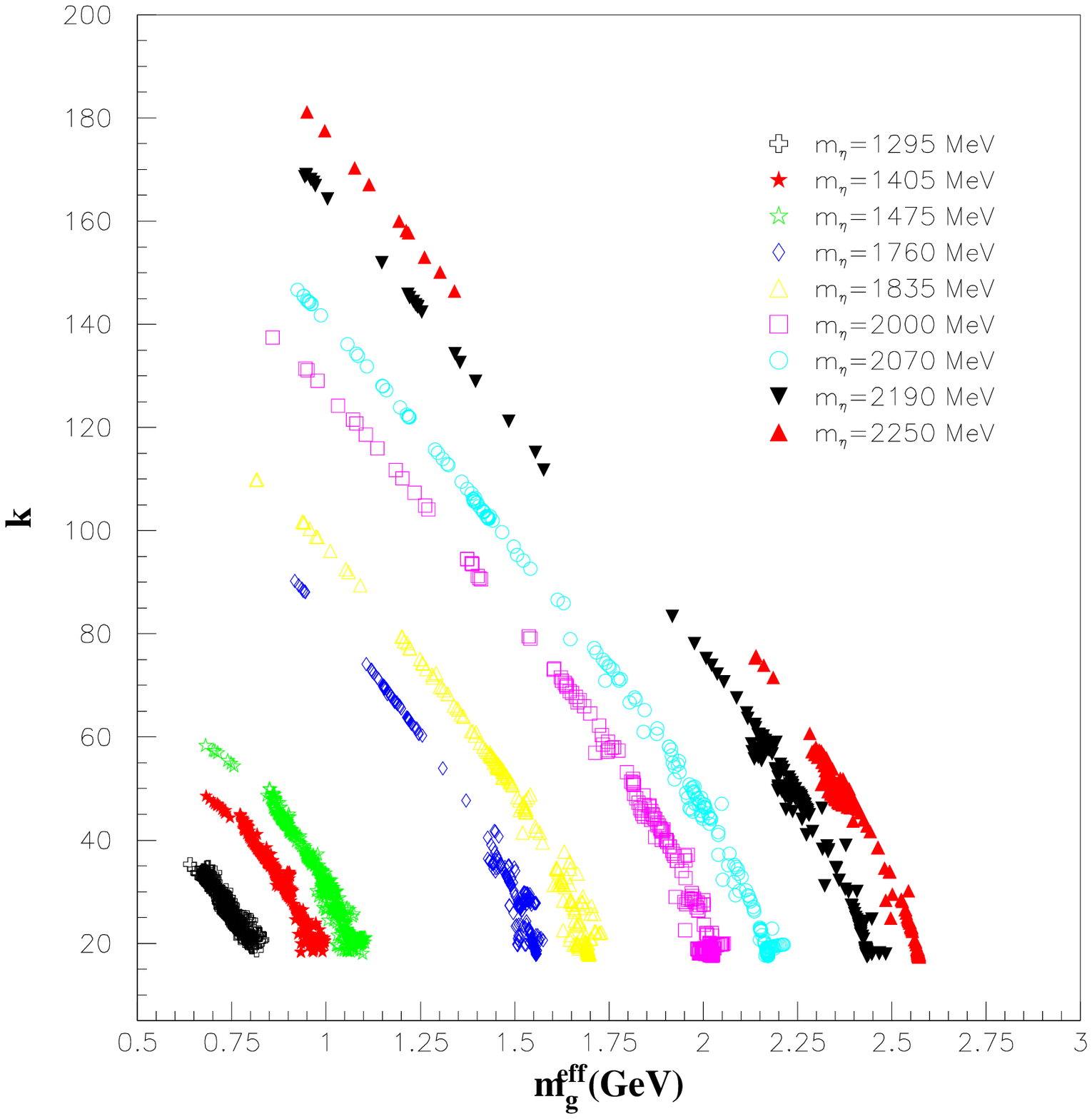}
}
\vskip -0.05cm \hskip 0.0 cm \textbf{( a ) } \hskip 6.6 cm \textbf{( b )}
\centerline{
\epsfxsize=6.5 cm \epsfysize=6.0cm \epsfbox{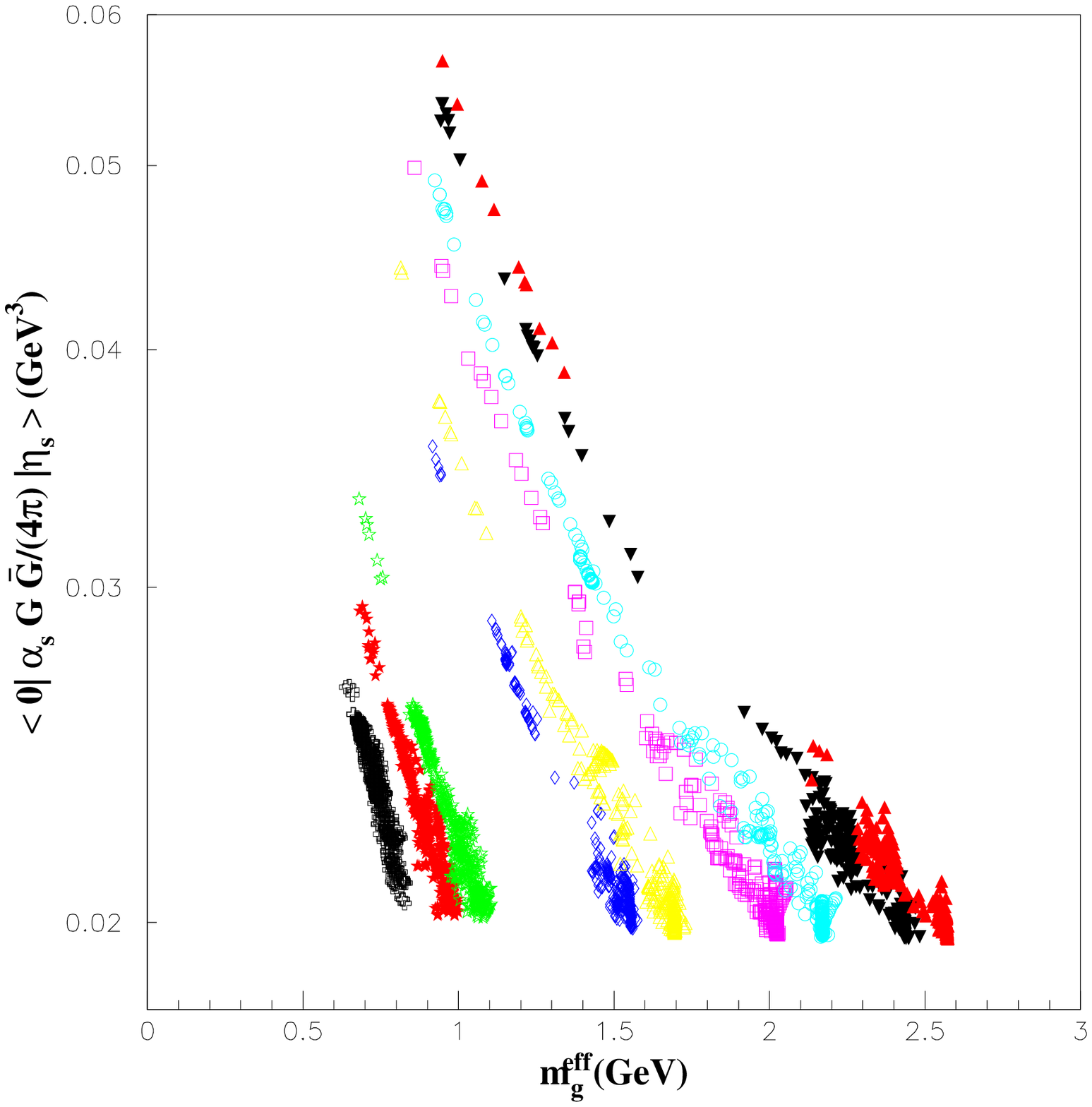}
\hspace*{0.2cm}
\epsfxsize=6.5 cm \epsfysize=6.0cm \epsfbox{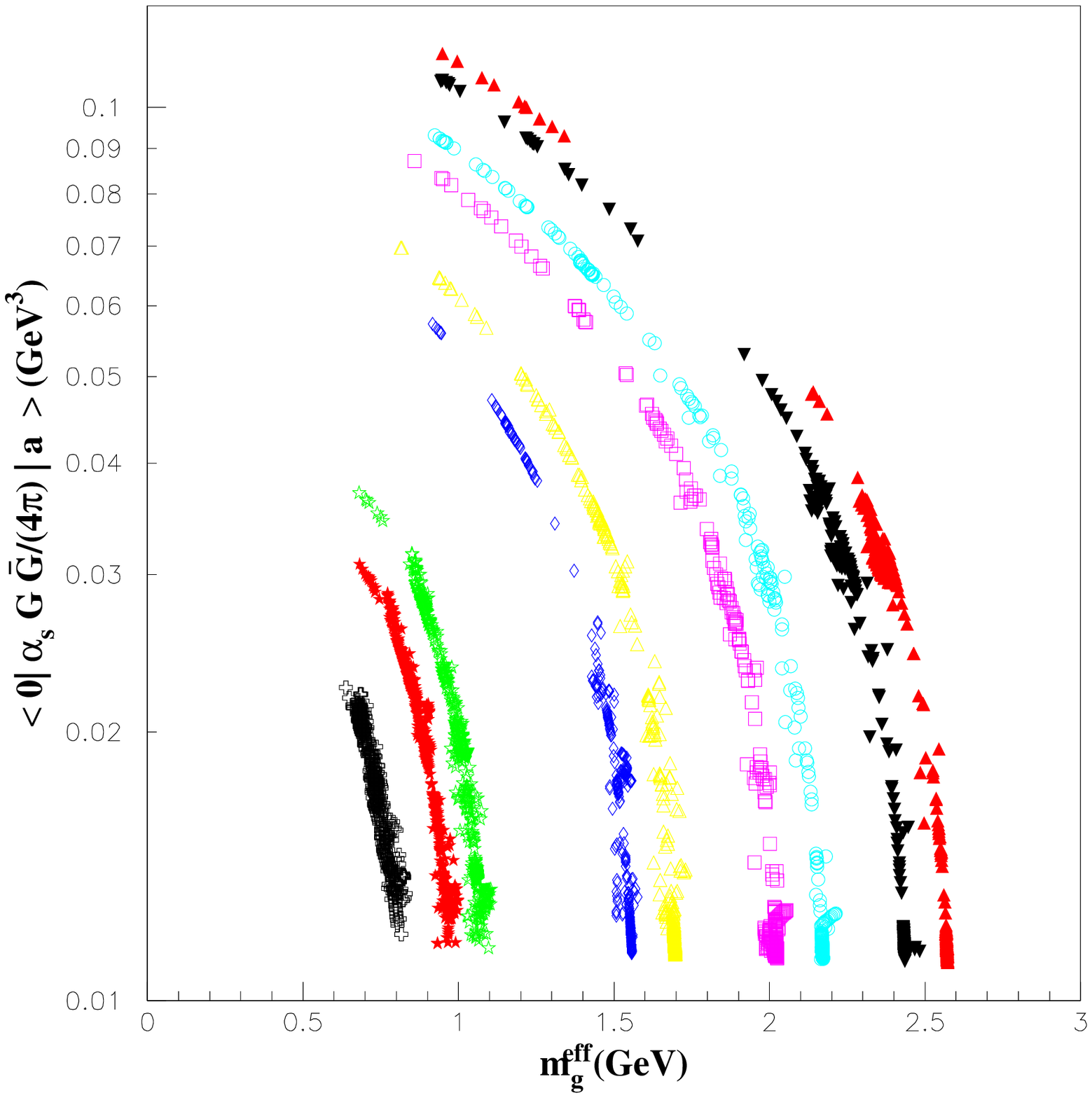}
}
\vskip -0.05cm \hskip 0.0 cm \textbf{( c ) } \hskip 6.6 cm \textbf{( d )}
\caption{\textit{The solutions for nine possible pseudoscalar glueball are demonstrated. The x-axis is for the effective glue mass. The y-axis is the product of $k$ and $v_Y$ for (a), $k$ for (b), $\langle 0 | \frac{\alpha_s}{4\pi}\,G\tilde G | \eta_s  \rangle$ for (c), and $\langle 0 | \frac{\alpha_s}{4\pi}\,G\tilde G | a  \rangle$ for (d). We only show solutions with $v_Y < 0.4$ GeV. The convention for marker in (c) and (d) is the same as in (a) and (b).}}
\label{fig-k-mge}
\end{figure}

Fig. (\ref{fig-k-mge}) is devoted to show the solutions in the parameter space for each case. Fig. (\ref{fig-k-mge}) demonstrates several facts: 1) Although the parameter $k$ can extend from $20$ to $180$ for these nine cases, the product of $k$ and $v_Y$ is situated in a quite narrow range (from $7$ to $18$), as shown in plot (a). Generally speaking, the small $v_Y$ corresponds to the large $k$, vice versa. This indicates that the term $k \, v_Y$ in the mass matrix given in Eq. (\ref{massmg08}) has more influence than the term $\frac{k}{v_Y}$ in determining the solutions. By using the relation between the effective field theory and the FKS method, $k v_Y$ is related with the values of $\langle 0 | \frac{\alpha_s}{4\pi}\,G\tilde G | \eta_q \rangle $ and $\langle 0 | \frac{\alpha_s}{4\pi}\,G\tilde G | \eta_s  \rangle$. Then from Fig.  (\ref{fig-k-mge} c), we can read out these transition elements: $\langle 0 | \frac{\alpha_s}{4\pi}\,G\tilde G | \eta_s  \rangle$ is within $(0.02, 0.06)$ GeV$^3$ and $\langle 0 | \frac{\alpha_s}{4\pi}\,G\tilde G | \eta_q \rangle$ is within $(0.04, 0.12)$ GeV$^3$, which is simply twice of $\langle 0 | \frac{\alpha_s}{4\pi}\,G\tilde G | \eta_s  \rangle$. These results are reasonable when compared with lattice results \cite{Chen:2005mg,Meyer:2008tr}. 2) When the product of $k$ and $v_Y$ is fixed, heavier glueball candidates have solutions with larger value of $m_g^{eff}$. Roughly speaking, the correlation between the product of $k$ and $v_Y$ and $m_g^{eff}$ is linear, quite similar to that between $k$ and $m_g^{eff}$. The value of $k$ is directly related with the transition element $\langle 0 | \frac{\alpha_s}{4\pi}\,G\tilde G | a  \rangle$, which is determined to be within the range $(0.01, 0.1)$ GeV$^3$, as demonstrated by (\ref{fig-k-mge}d). 3) A heavier glueball has a wider range of $k$ and $k \, v_Y$. 

Solutions in Fig. \ref{fig-k-mge}) demonstrates a lower bound for $m_g^{eff}$ with $m_g^{eff}>0.5$ GeV, which is larger than the lower bound determined from lattice and Dyson-Schwinger methods.
It is also worthy of mentioning that the information on the effective gluon mass can be helpful to rule out lots of solutions. For instance, solutions with $m_g^{eff}>1.2$ should be excluded if we treat the upper bound of $m_g^{eff}$ from lattice and Dyson-Schwinger methods seriously.

Below we calculate the branching fraction of pseudoscalar glueball candidates (which is denoted as $G$ here) decaying to two photons, decaying to two electrons, and to two muons. We assume that glueball only decays into these channel via its quark component \cite{dmli}. With this assumption, the ratio of the $G\to\gamma\gamma$ width over the $\pi^0\to\gamma\gamma$ one is expressed as
\begin{eqnarray}
\frac{\Gamma(G\to\gamma\gamma)}{\Gamma(\pi^0\to\gamma\gamma)}
&=&\frac{1}{9}\left(\frac{m_{gg}}{m_{\pi^0}}\right)^3\left(5\frac{f_\pi}{f_q}
V_{gq}+\sqrt{2}\frac{f_\pi}{f_s}V_{gs} \right)^2\;.
\label{R2}
\end{eqnarray}
For the $G\to\ell^+\ell^-$ decays, it is possible to determine their widths by using the available $\pi^0\to e^+e^-$ and $\eta\to\mu^+\mu^-$ data:
\begin{eqnarray}
\frac{\Gamma(G\to e^+e^-)}{\Gamma(\pi^0\to e^+e^-)}
&=&\frac{1}{9}\left(\frac{m_{gg}}{m_{\pi^0}}\right)^3\left(5\frac{f_\pi}{f_q}
V_{gq}+\sqrt{2}\frac{f_\pi}{f_s}V_{qs} \right)^2\;,
\nonumber\\
\frac{\Gamma(G\to\mu^+\mu^-)}{\Gamma(\eta\to\mu^+\mu^-)}
&=&\left(\frac{m_{gg}}{m_{\eta}}\right)^3\left(5\frac{f_\pi}{f_q}
V_{gq} +\sqrt{2}\frac{f_\pi}{f_s} V_{qs} \right)^2
\times \left[5\frac{f_\pi}{f_q}
V_{\eta q}
+ \sqrt{2}\frac{f_\pi}{f_s} V_{\eta s} \right]^{-2}\;.
\end{eqnarray}
These decay widths only depend on the mass of $0^{-+}$ glueball candidates, $m_{gg}$ and the corresponding eigenvectors.

\begin{figure}[t]
\centerline{
\epsfxsize=6.5 cm \epsfysize=6.0cm \epsfbox{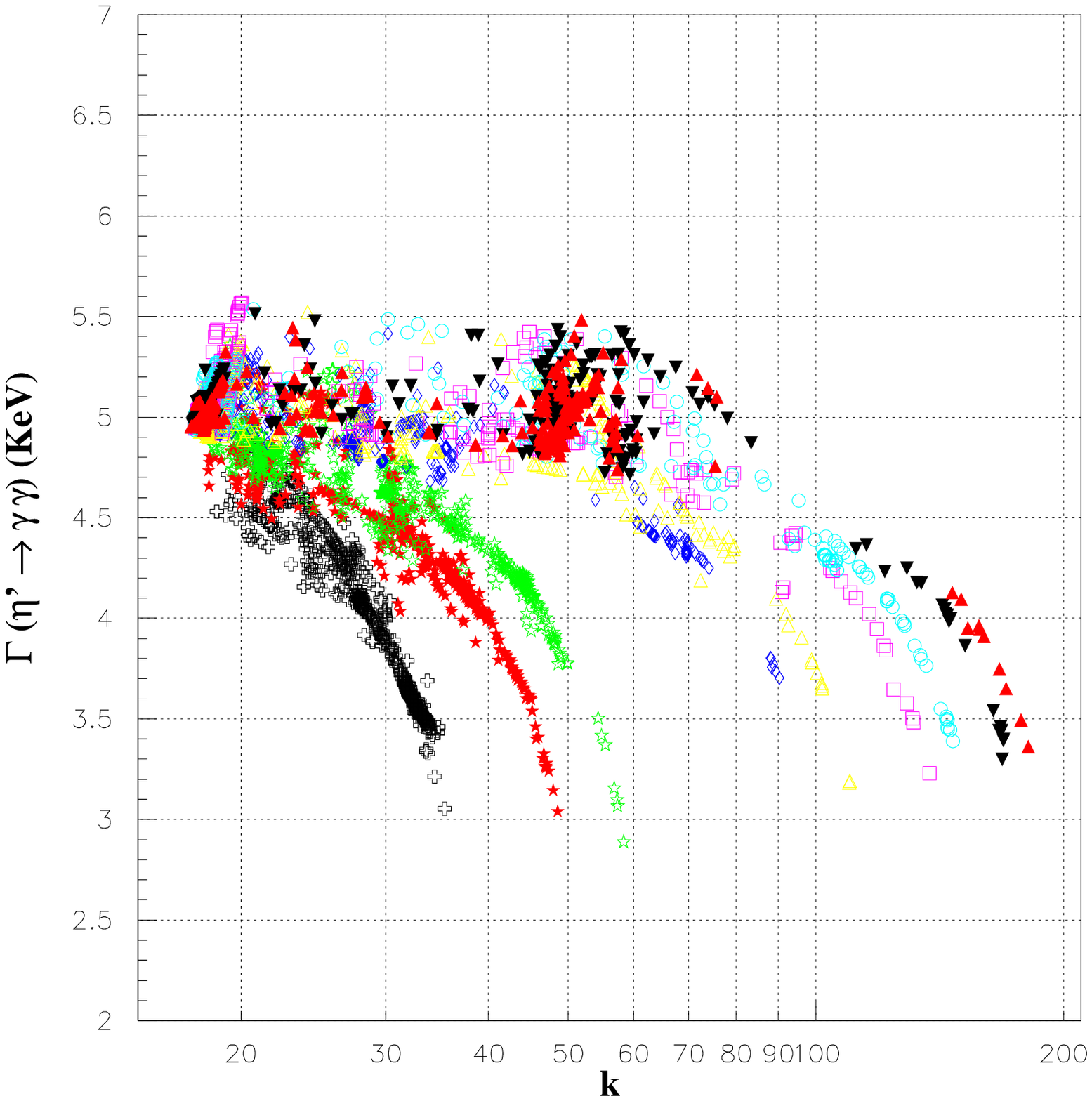}
\hspace*{0.2cm}
\epsfxsize=6.5 cm \epsfysize=6.0cm \epsfbox{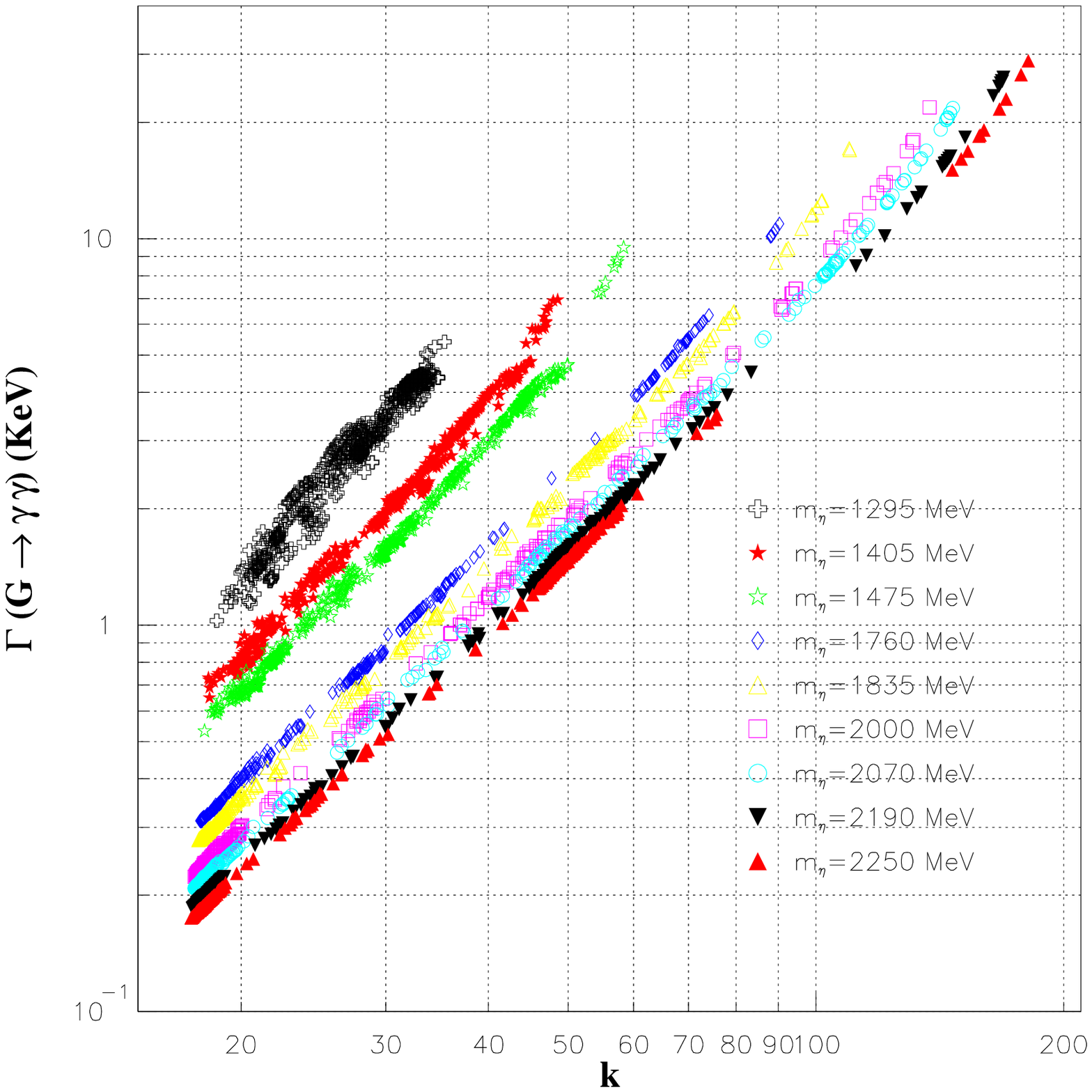}
}
\vskip -0.05cm \hskip 0.0 cm \textbf{( a ) } \hskip 6.6 cm \textbf{( b )}
\centerline{
\epsfxsize=6.5 cm \epsfysize=6.0cm \epsfbox{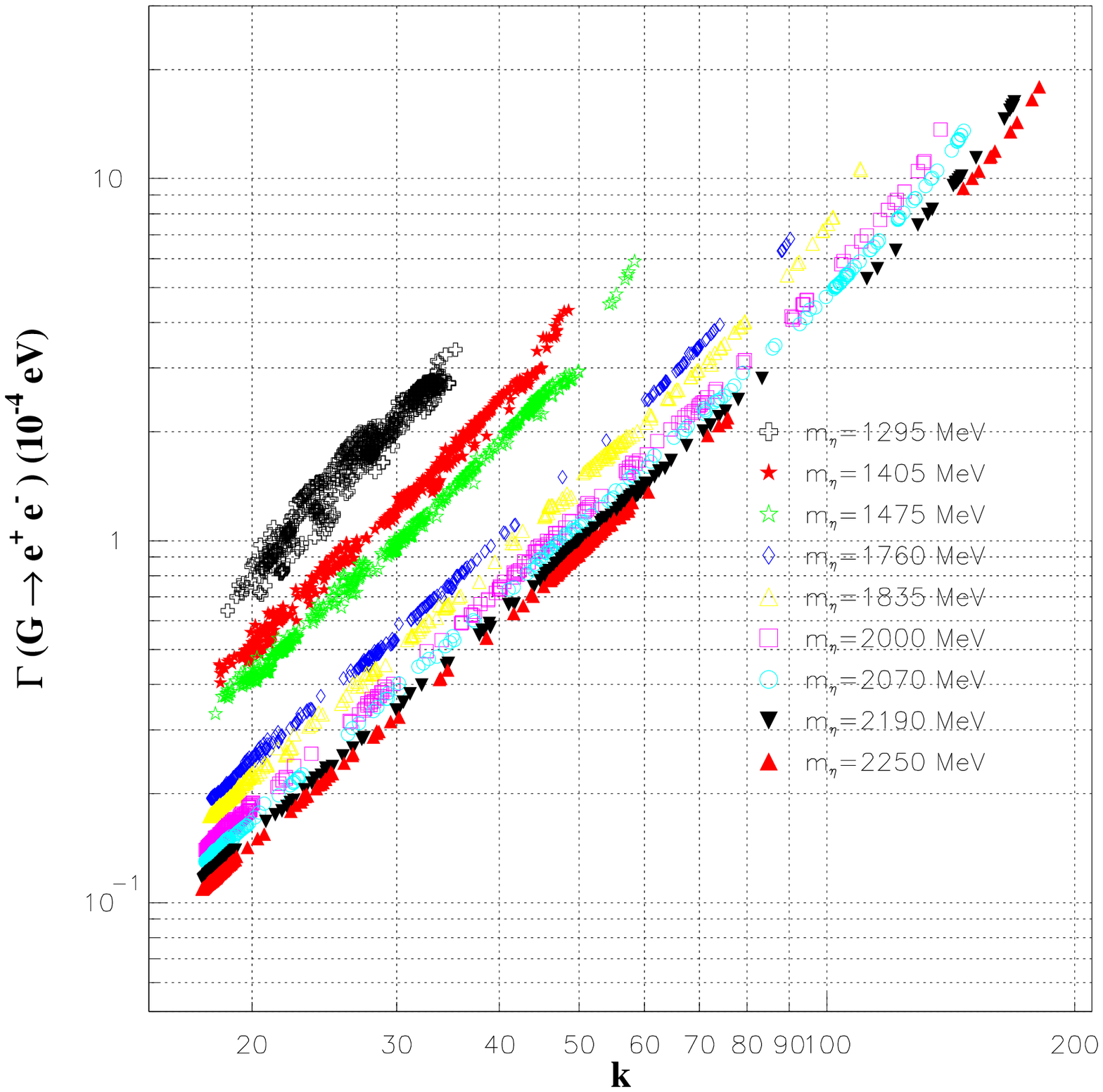}
\hspace*{0.2cm}
\epsfxsize=6.5 cm \epsfysize=6.0cm \epsfbox{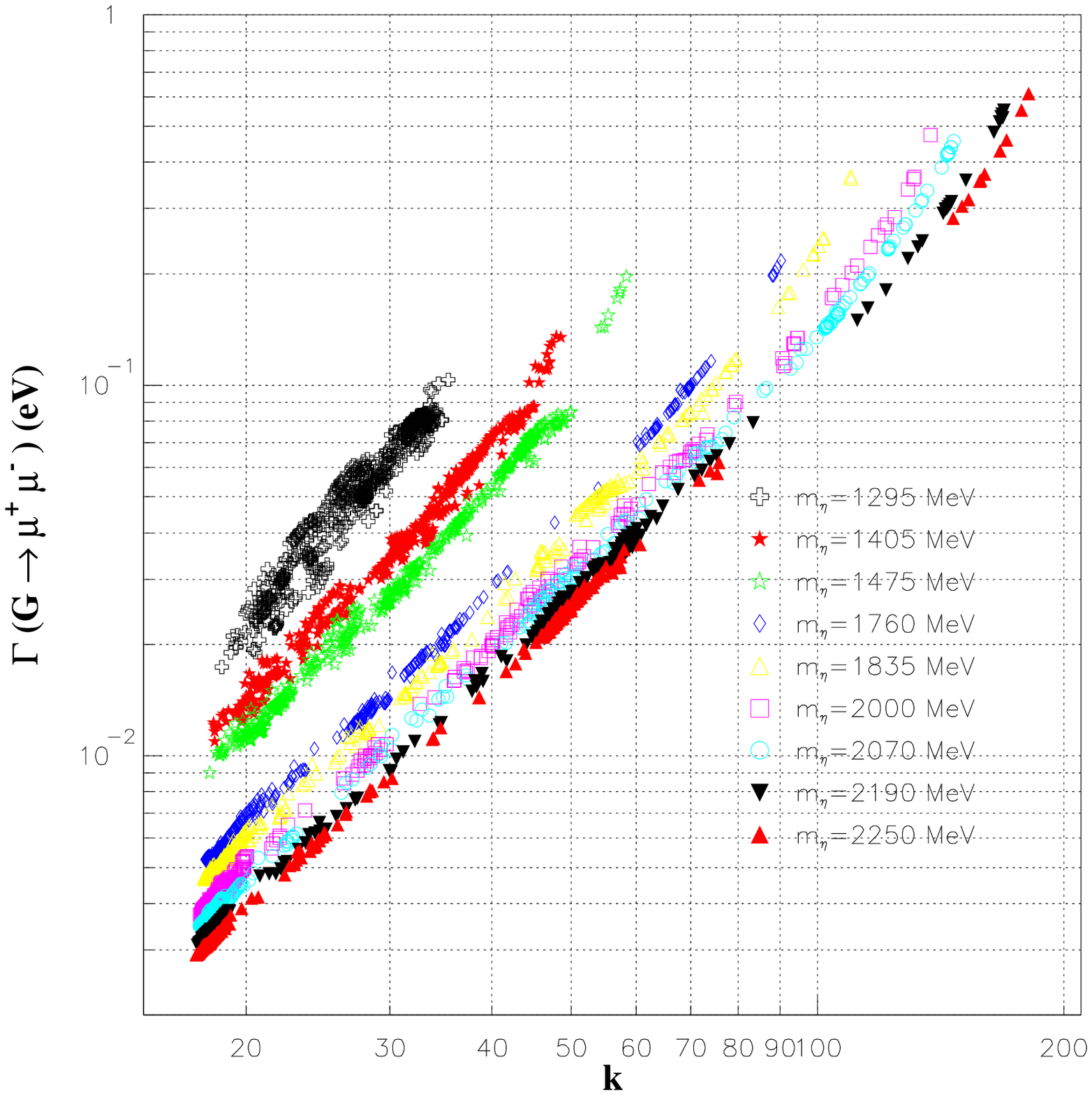}
}
\vskip -0.05cm \hskip 0.0 cm \textbf{( c ) } \hskip 6.6 cm \textbf{( d )}
\caption{\textit{The dependence of decay widths of glueball candidates on parameter $k$ are shown. In order to compare and contrast, we also show the decay width of $\eta^\prime$ to two photons. In plot (a), the convention of markers is the same as the rest of plots.}}
\label{fig-dwt}
\end{figure}

Fig. (\ref{fig-dwt}) is devoted to study these decay widths, which might be useful to identify glueball candidates from future experiments. We notice that $\Gamma( \eta \to \gamma \gamma)$ can change in the range $(0.36, 0.45)$ KeV (slightly lower than its experimental value $0.51$ KeV) \cite{Amsler:2008zz} and $\Gamma(\eta^\prime \to \gamma \gamma)$ can change in the range $(3, 5.5)$ KeV( its experimental value $4.28$ KeV is within this range \cite{Amsler:2008zz} ), both of which are not very sensitive to the change of $k$. However, as we can still read out from Fig.  (\ref{fig-dwt}a), for each case, when $k$ is large, $\Gamma(\eta^\prime \to \gamma \gamma)$ decreases slightly. The underlying reason is that when $k$ increase, more glueball component enters into $\eta^\prime$, which reduces the quark components in effect. 

While in each case, $\Gamma( G \rightarrow \gamma \gamma)$, $\Gamma( G \rightarrow e^+ e^-)$, and $\Gamma( G \rightarrow \mu^+ \mu^- )$ increase with the increase of $k$, as demonstrated in Fig.  (\ref{fig-dwt}b-\ref{fig-dwt}d). These results demonstrate that it is quite  a challenge to measure these processes from experiments. For each case, the relation between decay width and the parameter $k$ is simply linear, since large $k$ means that more quark states mix into the glueball state, as revealed by the mass matrixes Eqs. (\ref{massm},\ref{massmgqs},\ref{massmg08}). Another remarkable fact is that the heavier the glueball candidate, the wider its decay widths might expand in the model.

\begin{figure}[t]
\centerline{
\epsfxsize=6.5 cm \epsfysize=6.0cm \epsfbox{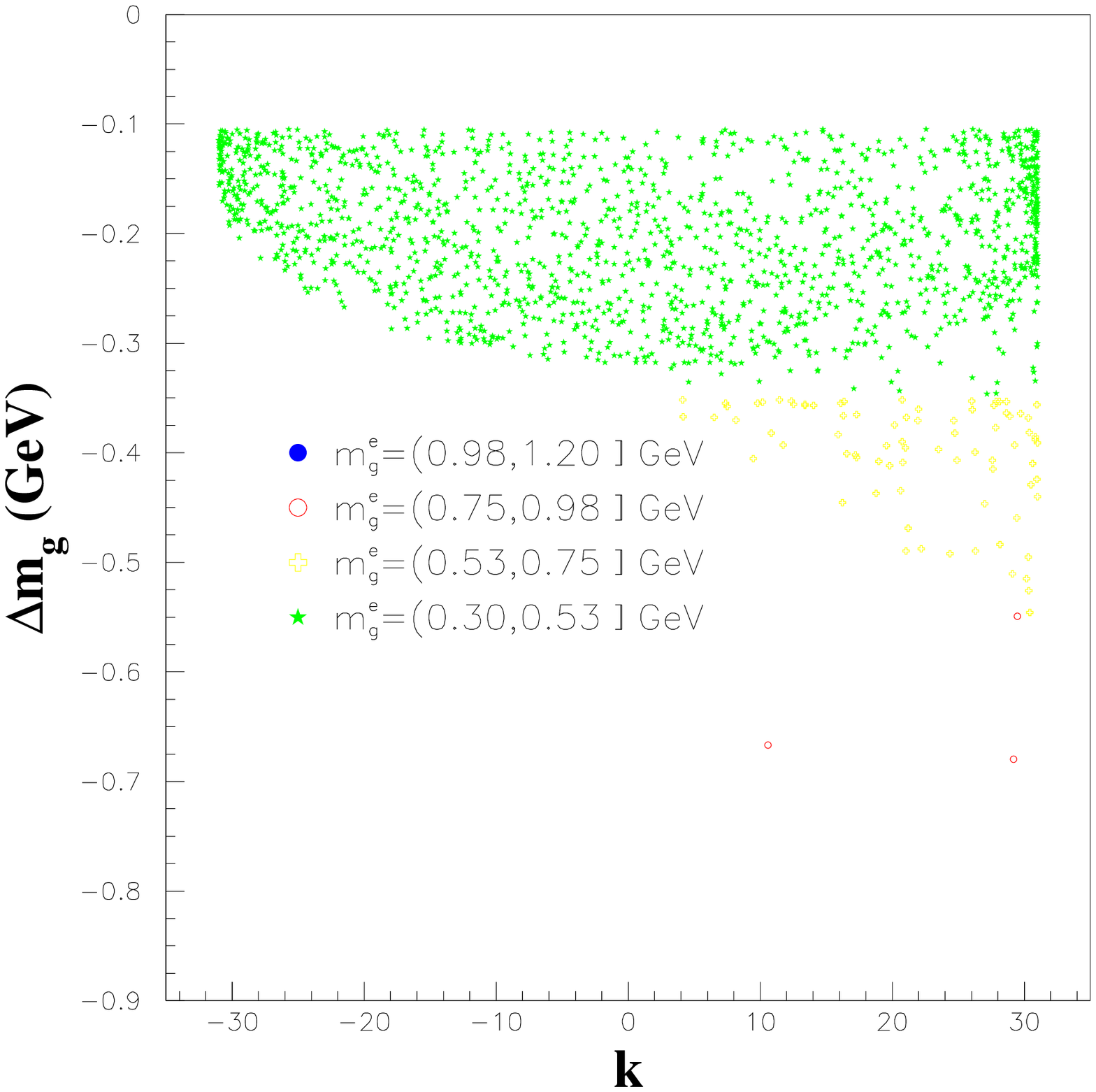}
\hspace*{0.4cm}
\epsfxsize=6.5 cm \epsfysize=6.0cm \epsfbox{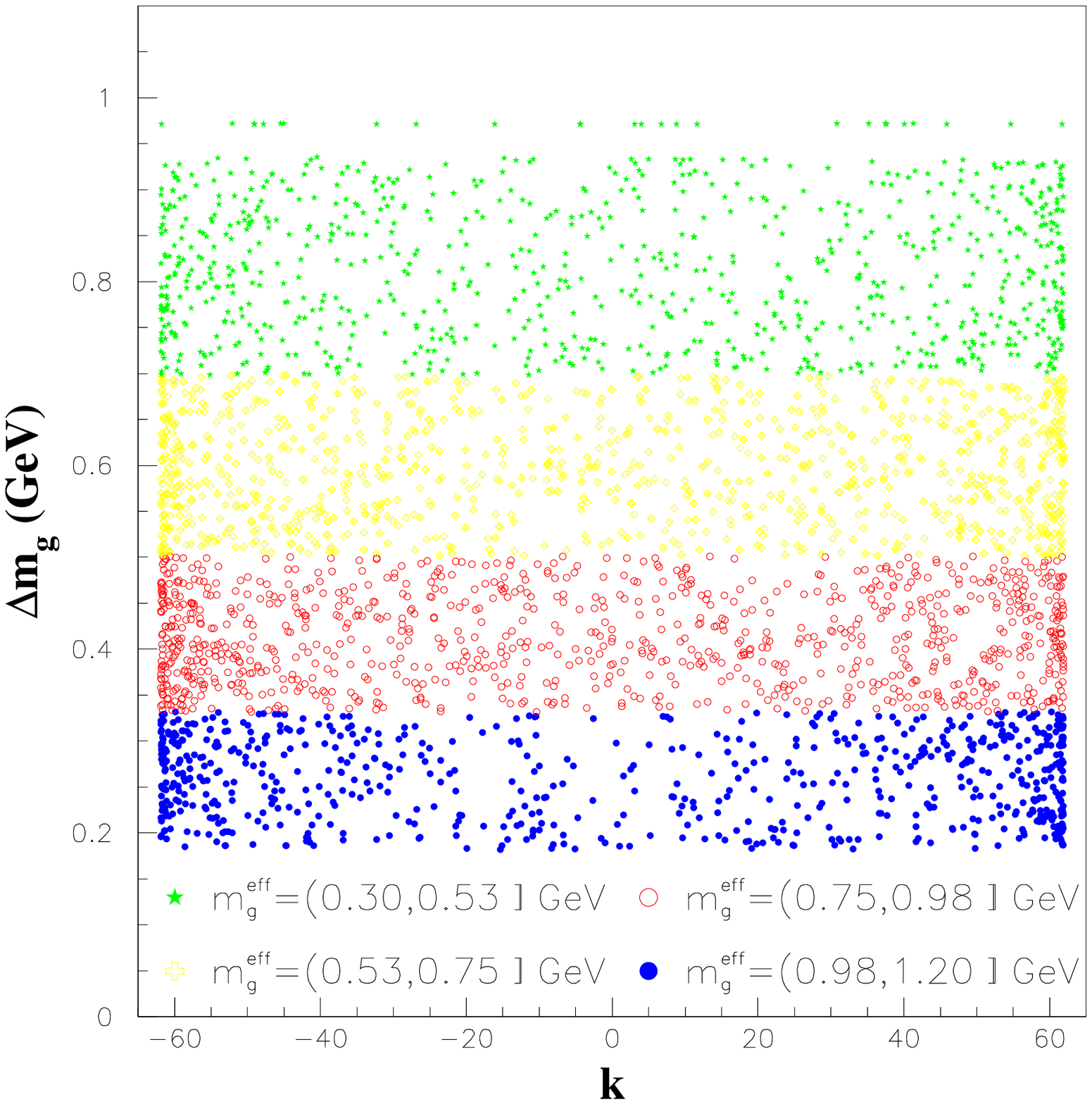}
}
\vskip -0.05cm \hskip 0.0 cm \textbf{( a ) } \hskip 6.6 cm \textbf{( b )}
\caption{\textit{Mass splitting between the scalar and pseudoscalar glueballs is demonstrated for two $0^{++}$ scenarios.}}
\label{fig-ms}
\end{figure}

It is also interesting to study how the mass splitting between the $0^{++}$ and $0^{-+}$ glueball. Fig. (\ref{fig-ms}) is devoted to study the mass splitting between the $0^{++}$ and $0^{-+}$ glueball states. We define $\Delta m_g = m_{0^{++}} - m_{0^{-+}}$. In the $0^{++}$ sector, the mass of glueball candidate is fixed by assumption. In the $0^{-+}$ sector, we identify the eigenstate with the largest $V_{ig}^2, i=\eta, \eta^\prime, G$ as the glueball candidate. From Fig. (\ref{fig-ms}), we observe that in both cases, the mass splitting is simply controlled by the effective gluon mass. In the first case, the larger effective mass of gluon corresponds to a larger mass splitting and the $0^{-+}$ glueball state is heavier than the $0^{++}$ state which is similar to the lattice results \cite{Chen:2005mg}, while in the second case, the situation is opposite. Therefore, if the mass of the glueball candidate is large than $1295$ MeV, it is difficult to realize in the first scenario. While for the second scenario, only those states with mass smaller than $1760$ MeV can be realized. Another interesting fact as shown in Fig. (\ref{fig-ms}) is that the mass splitting is almost independent on the parameter $k$.

\section{Discussions and Conclusions}

In this work, we use the effective Lagrangian derived from the instanton effects to study the pseudoscalar glueball candidates. At the leading approximation, we observe that the results obtained in this model are consistent with other methods, like FKS method and the lattice method, which indicates that this effective Lagrangian captures the most important features of QCD.

We notice that the $0^{++}$ and $0^{-+}$ sectors favor different regions of the parameter space, therefore the physics regions of parameters for both sectors are constrained. Without considering the Regge excitations, we find that the parameter space for the first scenario given in Table \ref{table-cases} for the $0^{++}$ sector can not be consistent with the known masses of pseudoscalar glueball candidates while the second scenario for the $0^{++}$ sector can. In the second scenario given in Table \ref{table-cases} for the $0^{++}$ sector, the pseudoscalar glueball candidates cannot be heavier than $\eta(1760)$, therefore only three candidates, i.e. $\eta(1295)$, $\eta(1405)$, and $\eta(1475)$, are possible pseudoscalar glueball candidates. 

To study the parameter space of $m_g^{eff}$, we have assumed that $D_0 = B_0$. Currently, we do not know the exact relation between $D$ and $m_g^{eff}$ from the underlying QCD theory. If $D_0$ is much smaller than $B_0$, then the parameter $m_g^{eff}$ can not be constrained too much. While if assuming $D_0 = 3 B_0$, we notice that the unitarity conditions rule out all points in the first scenario given in Table \ref{table-cases} of $0^{++}$ sector and only few points survive for the second scenario given in Table \ref{table-cases}.

We have combined both $0^{++}$ and $0^{-+}$ sectors to study the constraints on the parameter space. In the $0^{++}$ sector, we only consider the mixing between the singlet $\sigma^0$ and the $0^{++}$ glueball. In a more realistic treatment, we should take into account the realization of the experimental $0^{++}$ spectra. Then it is necessary to extend the Lagrangian by including the tetraquark state \cite{Black:1998wt,Hooft:2008we}. Furthermore, we have utilized the nonlinear realization formalism to treat the $0^{-+}$ sector. In order to take into account the interaction between $0^{++}$ and $0^{-+}$ sectors, maybe it's better to use the linear realization form to parameterize both sectors, as demonstrated by \cite{Minkowski:1998mf,'tHooft:1999jc}. However, the main conclusions given here might not be changed.

We have found the correspondence between the FKS method and the effective field theory method at the leading order. In the effective field theory method, the mass matrix is automatically symmetric by requiring the Lagrangian is Hermitian. While in the FKS method beyond the leading order, the mass matrix is not necessarily symmetric. In order to have a symmetric mass matrix, we must invoke an asymmetric decay constant matrix (\ref{eq:j}), which is difficult to find its counterpart in the effective field theory method.

The 4D effective field theory can be extended to accommodate higher excitations of by putting the masses of pseudoscalar on the Regge lines. Interactions can be introduced by invoking the assumption that there exists a global symmetry for each radial excitations multiplet, as suggested in \cite{Glozman:2007ek}. An alternative way is to use the AdS/QCD in the soft-wall model \cite{KKSS2006,Katz:2007tf}, which can accommodate the Regge trajectories in a more convenient way. We will extend our analysis on pseudoscalar glueball candidates to this model in our future work.

\section{Acknowledgment}
The work of M.H. is supported by
CAS program "Outstanding young scientists abroad brought-in", the CAS key project under grant No. KJCX3-SYW-N2, and NSFC under grant No. 10875134 and No. 10735040. Q.S.Y. thanks the hospitality of TPCSF during his stay when this work was initiated and the hospitality of the Key Laboratory of Particle Astrophysics, IHEP, CAS in the final stage of the work.

\section{Appendix}
\setcounter{equation}{0}
\renewcommand{\theequation}{\arabic{section}.\arabic{equation}}%
\subsection{The symmetric FKS formalism with glueball}
We follow the convention of \cite{Cheng:2008ss} and list the FKS formalism by including a glueball candidate. We extend the asymmetric mass formalism given in \cite{Cheng:2008ss} to a symmetric $3\times 3$ mass matrix.

The basic difference in our formalism is that we can define a vector current $K^{\mu}$ of gluon fields
\bea
K^{\mu} &=& \frac{ \alpha_s}{4 \pi} \epsilon^{\mu\nu\alpha\beta} A_\nu^a \left ( \partial_\alpha A_\beta^a + \frac{g}{3} f^{abc} A^b_\alpha A^c_\beta \right )\,.
\eea
At the same time, for the quark sector, we can define two axial-vector currents, which read
\bea
J^\mu_{q,5} &=& \bar q\gamma^\mu\gamma_5 q\,, \\
J^\mu_{s,5} &=& \bar s \gamma^\mu\gamma_5 s\,. 
\eea
Therefore, beside the standard FKS decay constants (vacuum expectation values) $v_q$ and $v_s$ given below
\begin{eqnarray}
   \langle 0|J_{q,5}^\mu|\eta_q(P)\rangle
   &=& -\frac{i}{\sqrt2}\,v_q\,P^\mu \;,\nonumber \\
   \langle 0|J^\mu_{s,5}|\eta_s(P)\rangle
   &=& -i v_s\,P^\mu \;,\label{deffq}
\end{eqnarray}
we have an extra definition
\bea
\langle 0| K^\mu |a(P)\rangle
   &=& -\frac{i}{ \sqrt{3}}\,v_Y\,P^\mu \label{deffy}\;,
\eea
where $P$ is the momentum defined in 4D. To include the OZI rule violation effects, we introduce the following transition elements:
\bea
\langle 0| K^\mu |\eta_q (P)\rangle
   &=& -\frac{i}{ \sqrt{3}}\,f^g_q\,P^\mu \;,\nonumber \\
\langle 0| K^\mu |\eta_s(P)\rangle
   &=& -\frac{i}{ \sqrt{3}}\,f^g_s\,P^\mu \;,\nonumber \\
   \langle 0|J^\mu_{q,5}|\eta_s(P)\rangle
   &=& -\frac{i}{\sqrt2}\,f^q_s\,P^\mu \;,\nonumber \\
   \langle 0|J^\mu_{q,5}|a(P)\rangle
   &=& -\frac{i}{\sqrt2}\,f^q_a\,P^\mu \;,\nonumber \\
   \langle 0|J^\mu_{s,5}|\eta_q(P)\rangle
   &=& -i f^s_q\,P^\mu \;,\nonumber \\
   \langle 0|J^\mu_{s,5}| a(P)\rangle
   &=& -i f^s_a\,P^\mu \;.\label{vozi}
\eea
The $f^i_j$ denotes the decay width of state $j$ to vacuum via operator $i$. At leading approximation, $f^i_j$ vanishes.

The physical states, the psuedo-scalar glueball state $G$, $\eta^\prime$, and $\eta$ are the mixtures of three basis ingredients $a$, $\eta_q$, and $\eta_s$, which can be represented as
\begin{equation}\label{qs}
   \left( \begin{array}{c}
    | G  \rangle \\ | \eta' \rangle \\ |\eta \rangle
   \end{array} \right)
   = U \cdot 
   \left( \begin{array}{c}
    |a \rangle \\ | \eta_q \rangle \\ \eta_s \rangle 
   \end{array} \right) \;,
\end{equation}
where $U$ is an orthogonal normalized matrix. Then relation between the decay width of physical states and of basis state is given as
\begin{equation}\label{qs}
F   = U \cdot 
   \left( \begin{array}{ccc}
    \frac{1}{\sqrt{3}} v_Y & \frac{1}{\sqrt{2}}f^q_a & f^s_a \\ 
    \frac{1}{\sqrt{3}}f^g_{\eta_q} & \frac{1}{\sqrt{2}}v_q & f^s_{\eta_q} \\
    \frac{1}{\sqrt{3}}f^g_{\eta_s} & \frac{1}{\sqrt{2}}f^q_{\eta_s} & v_s     
   \end{array} \right) \;, 
\end{equation}
where $F$ is defined as
\bea
F =   \left( \begin{array}{c c c }
    f^g_G & f^q_G & f^s_G \\ 
    f^g_{\eta^\prime} & f^q_{\eta^\prime} & f^s_{\eta^\prime} \\
    f^g_{\eta} & f^q_{\eta} & f^s_{\eta}     
   \end{array} \right) \,.
\eea

Then we can have one identity and the two anomalous Ward identity, which read 
\bea
\partial_\mu K^\mu &=& \frac{\alpha_s}{4\pi}\,G_{\mu\nu}\,\widetilde{G}^{\mu\nu}\;, \label{id1}\\
   \partial_\mu J^\mu_{q,5} &=& 2im_q\,\bar q\gamma_5 q
   +\frac{\alpha_s}{4\pi}\,G_{\mu\nu}\,\widetilde{G}^{\mu\nu}\;,\\
\partial_\mu J^\mu_{s,5}&=& 2im_s\,\bar s\gamma_5s
   +\frac{\alpha_s}{4\pi}\,G_{\mu\nu}\,\widetilde{G}^{\mu\nu} \label{id3}\;.
\eea
By using $\langle 0 |$ and physical states to sandwich these three equations, we arrive the following equation:
\begin{equation}\label{mass}
   \left( \begin{array}{c c c }
   \langle 0|\partial_\mu K^\mu | G \rangle &  \langle 0|\partial_\mu J_{q,5}^\mu | G \rangle &  \langle 0|\partial_\mu J_{s,5}^\mu | G \rangle \\ 
   \langle 0|\partial_\mu K^\mu | \eta^\prime \rangle & \langle 0|\partial_\mu J_{q,5}^\mu | \eta^\prime \rangle&\langle 0|\partial_\mu J_{s,5}^\mu | \eta^\prime \rangle  \\
   \langle 0|\partial_\mu K^\mu | \eta \rangle & \langle 0|\partial_\mu J_{q,5}^\mu | \eta \rangle & \langle 0|\partial_\mu J_{s,5}^\mu | \eta \rangle      
   \end{array} \right)
   = M^2_D \cdot F \,,
\end{equation}
where $M_D^2$ is given as
\bea
M^2_D =   \left( \begin{array}{c c c }
    M_G^2 &  &   \\ 
     & M^2_{\eta^\prime} &   \\
      &   & M^2_{\eta}     
   \end{array} \right) \,.
\eea
To obtain this equation, we just use the definition given in Eqs. (\ref{deffq}-\ref{vozi}).

According to Eq. (\ref{qs}), we can also express right hand side of Eq. (\ref{mass}) as
\begin{equation}\label{relation}
   \left( \begin{array}{c c c }
   \langle 0|\partial_\mu K^\mu | G \rangle &  \langle 0|\partial_\mu J_{q,5}^\mu | G \rangle &  \langle 0|\partial_\mu J_{s,5}^\mu | G \rangle \\ 
   \langle 0|\partial_\mu K^\mu | \eta^\prime \rangle & \langle 0|\partial_\mu J_{q,5}^\mu | \eta^\prime \rangle&\langle 0|\partial_\mu J_{s,5}^\mu | \eta^\prime \rangle  \\
   \langle 0|\partial_\mu K^\mu | \eta \rangle & \langle 0|\partial_\mu J_{q,5}^\mu | \eta \rangle & \langle 0|\partial_\mu J_{s,5}^\mu | \eta \rangle      
   \end{array} \right)
   = U \cdot  \left( \begin{array}{c c c }
   \langle 0|\partial_\mu K^\mu | a \rangle &  \langle 0|\partial_\mu J_{q,5}^\mu | a \rangle &  \langle 0|\partial_\mu J_{s,5}^\mu | a \rangle \\ 
   \langle 0|\partial_\mu K^\mu | \eta_q \rangle & \langle 0|\partial_\mu J_{q,5}^\mu | \eta_q \rangle&\langle 0|\partial_\mu J_{s,5}^\mu | \eta_q \rangle  \\
   \langle 0|\partial_\mu K^\mu | \eta_s \rangle & \langle 0|\partial_\mu J_{q,5}^\mu | \eta_s \rangle & \langle 0|\partial_\mu J_{s,5}^\mu | \eta_s \rangle      
   \end{array} \right)
\,.
\end{equation}
By using Eqs. (\ref{id1}-\ref{id3}), the last term in the right hand side of 
Eq. (\ref{relation}) can be represented as
\bea\label{fks}
\left( \begin{array}{c c c }
   \langle 0|\partial_\mu K^\mu | a \rangle & \frac{v_Y}{\sqrt{3}} m^2_{qa} +  \langle 0|\partial_\mu K^\mu | a \rangle & \frac{v_Y}{\sqrt{3}}  m^2_{sa} + \langle 0|\partial_\mu K^\mu | a \rangle \\ 
   \langle 0|\partial_\mu K^\mu | \eta_q \rangle &  \frac{v_q}{\sqrt{2}}  m^2_{qq} + \langle 0| \partial_\mu K^\mu | \eta_q \rangle&  \frac{v_q}{\sqrt{2}}  m^2_{sq} + \langle 0|\partial_\mu K^\mu | \eta_q \rangle  \\
   \langle 0|\partial_\mu K^\mu | \eta_s \rangle & v_s m^2_{qs} + \langle 0|\partial_\mu K^\mu | \eta_s \rangle & v_s m^2_{ss} + \langle 0| \partial_\mu K^\mu | \eta_s \rangle      
   \end{array} \right)
   &=& M^2_{gqs} \cdot F^\prime \,,
\eea
with 
\bea
M_{gqs}^2 &=&  \left( \begin{array}{c c c }
  \frac{\sqrt{3}}{v_Y}  \langle 0|\partial_\mu K^\mu | a \rangle & m^2_{qa} + \frac{\sqrt{3}}{v_Y} \langle 0|\partial_\mu K^\mu | a \rangle &  m^2_{sa} +  \frac{\sqrt{3}}{v_Y} \langle 0|\partial_\mu K^\mu | a \rangle \\ 
  \frac{\sqrt{2}}{v_q} \langle 0|\partial_\mu K^\mu | \eta_q \rangle &    m^2_{qq} + \frac{\sqrt{2}}{v_q} \langle 0| \partial_\mu K^\mu | \eta_q \rangle& m^2_{sq} + \frac{\sqrt{2}}{v_q} \langle 0|\partial_\mu K^\mu | \eta_q \rangle  \\
  \frac{1}{v_s} \langle 0|\partial_\mu K^\mu | \eta_s \rangle & m^2_{qs} +  \frac{1}{v_s}  \langle 0|\partial_\mu K^\mu | \eta_s \rangle &  m^2_{ss} +  \frac{1}{v_s}  \langle 0| \partial_\mu K^\mu | \eta_s \rangle      
   \end{array} \right)\,, \\
F^\prime &=& \left( \begin{array}{c c c }
    \frac{1}{\sqrt{3}} v_Y &  &   \\ 
     & \frac{1}{\sqrt{2}} v_q &   \\
      &   & v_s     
   \end{array} \right)\,.
\eea
The quantities $m^2_{ij}$ are defined as
\bea
m^2_{qa, qq, qs} &=& \frac{\sqrt{2}}{v_q} \langle 0| i m_u {\bar u} \gamma_5 u  + i m_d {\bar d} \gamma_5 d | a, \eta_q, \eta_s \rangle\,,\\
m^2_{sa, sq, ss} &=& \frac{1}{v_s} \langle 0| i m_s {\bar s} \gamma_5 s  |a,  \eta_q, \eta_s \rangle\,,
\eea

Combining Eqs. (\ref{qs}, \ref{mass}, \ref{relation}, \ref{fks}), we arrive at the symmetric mass matrix which reads as
\bea\label{eq:sym}
M_{gqs}^2 \cdot J^{-1} &=& U^{\dagger} \cdot M^2_D \cdot U\,,
\eea
where $J$ is defined as
\begin{eqnarray}
\label{eq:j}
J = \left ( \begin{array}{ccc}
1 &  \frac{f^q_a}{v_q} &  \frac{f^s_a}{v_s} \\
 \frac{f^g_q}{v_Y} & 1 &  \frac{f^s_q}{v_s}\\
 \frac{f^g_s}{v_Y} &  \frac{f^q_s}{v_q} & 1 \\
\end{array} \right )\,.
\end{eqnarray} 
At the leading order approximation, $J^{-1}$ is an identity matrix. Therefore, we can match the left side with our mass matrix determined in the effective field theory method given in Eq. (\ref{massm}).

Obviously, in the general case, the mass matrix $M_{gqs}^2$ is asymmetric. In order to have a symmetry mass matrix, the asymmetric matrix $J$ is needed, which corresponds to take into account the higher order corrections.

\end{document}